\documentstyle[twoside,epsfig,psfig,axodraw]{article} 
 
\catcode`\@=11 
\long\def\@makefntext#1{ 
\protect\noindent \hbox to 3.2pt {\hskip-.9pt   
$^{{\eightrm\@thefnmark}}$\hfil}#1\hfill}               
 
\def\thefootnote{\fnsymbol{footnote}} 
\def\@makefnmark{\hbox to 0pt{$^{\@thefnmark}$\hss}}    
         
\def\ps@myheadings{\let\@mkboth\@gobbletwo 
\def\@oddhead{\hbox{} 
\rightmark\hfil\eightrm\thepage}    
\def\@oddfoot{}\def\@evenhead{\eightrm\thepage\hfil 
\leftmark\hbox{}}\def\@evenfoot{} 
\def\sectionmark##1{}\def\subsectionmark##1{}} 
 
\oddsidemargin=\evensidemargin 
\renewcommand{\thefootnote}{\fnsymbol{footnote}}

\newcounter{sectionc}\newcounter{subsectionc}\newcounter{subsubsectionc} 
\renewcommand{\section}[1] {\vspace{12pt}\addtocounter{sectionc}{1}  
\setcounter{subsectionc}{0}\setcounter{subsubsectionc}{0}\noindent  
        {\tenbf\thesectionc. #1}\par\vspace{5pt}} 
\renewcommand{\subsection}[1] {\vspace{12pt}\addtocounter{subsectionc}{1}  
        \setcounter{subsubsectionc}{0}\noindent  
        {\bf\thesectionc.\thesubsectionc. {\kern1pt \bfit #1}}\par\vspace{5pt}} 
\renewcommand{\subsubsection}[1] {\vspace{12pt}\addtocounter{subsubsectionc}{1} 
        \noindent{\tenrm\thesectionc.\thesubsectionc.\thesubsubsectionc. 
        {\kern1pt \tenit #1}}\par\vspace{5pt}} 
\newcommand{\nonumsection}[1] {\vspace{12pt}\noindent{\tenbf #1} 
        \par\vspace{5pt}} 
 
\newcounter{appendixc} 
\newcounter{subappendixc}[appendixc] 
\newcounter{subsubappendixc}[subappendixc] 
\renewcommand{\thesubappendixc}{\Alph{appendixc}.\arabic{subappendixc}} 
\renewcommand{\thesubsubappendixc} 
        {\Alph{appendixc}.\arabic{subappendixc}.\arabic{subsubappendixc}} 
 
\renewcommand{\appendix}[1] {\vspace{12pt} 
        \stepcounter{appendixc} 
        \setcounter{figure}{0} 
        \setcounter{table}{0} 
        \setcounter{lemma}{0} 
        \setcounter{theorem}{0} 
        \setcounter{corollary}{0} 
        \setcounter{definition}{0} 
        \setcounter{equation}{0} 
        \renewcommand{\thefigure}{\Alph{appendixc}.\arabic{figure}} 
        \renewcommand{\thetable}{\Alph{appendixc}.\arabic{table}} 
        \renewcommand{\theappendixc}{\Alph{appendixc}} 
        \renewcommand{\thelemma}{\Alph{appendixc}.\arabic{lemma}} 
        \renewcommand{\thetheorem}{\Alph{appendixc}.\arabic{theorem}} 
        \renewcommand{\thedefinition}{\Alph{appendixc}.\arabic{definition}} 
        \renewcommand{\thecorollary}{\Alph{appendixc}.\arabic{corollary}} 
        \renewcommand{\theequation}{\Alph{appendixc}.\arabic{equation}} 
        \noindent{\tenbf Appendix \theappendixc #1}\par\vspace{5pt}} 
\newcommand{\subappendix}[1] {\vspace{12pt} 
        \refstepcounter{subappendixc} 
        \noindent{\bf Appendix \thesubappendixc. {\kern1pt \bfit #1}} 
        \par\vspace{5pt}} 
\newcommand{\subsubappendix}[1] {\vspace{12pt} 
        \refstepcounter{subsubappendixc} 
        \noindent{\rm Appendix \thesubsubappendixc. {\kern1pt \tenit #1}} 
        \par\vspace{5pt}} 
 
\newcommand{\textlineskip}{\baselineskip=13pt} 
\newcommand{\smalllineskip}{\baselineskip=10pt} 
 
\def\eightcirc{ 
\begin{picture}(0,0) 
\put(4.4,1.8){\circle{6.5}} 
\end{picture}} 
\def\eightcopyright{\eightcirc\kern2.7pt\hbox{\eightrm c}}  
 
\newcommand{\copyrightheading}[1] 
        {\vspace*{-2.5cm}\smalllineskip{\flushleft 
        {\footnotesize International Journal of Modern Physics A, #1}\\ 
        {\footnotesize $\eightcopyright$\, World Scientific Publishing 
         Company}\\ 
         }} 
 
\newcommand{\publisher}[2]{{\begin{center}\footnotesize\smalllineskip  
        Received #1\\ 
        Revised #2 
        \end{center} 
        }} 
 
\def\abstracts#1#2#3{{ 
        \centering{\begin{minipage}{4.5in}\baselineskip=10pt\footnotesize 
        \parindent=0pt #1\par  
        \parindent=15pt #2\par 
        \parindent=15pt #3 
        \end{minipage}}\par}}  
 
\renewenvironment{thebibliography}[1] 
        {\frenchspacing 
         \ninerm\baselineskip=11pt 
         \begin{list}{\arabic{enumi}.} 
        {\usecounter{enumi}\setlength{\parsep}{0pt} 
         \setlength{\leftmargin 12.7pt}{\rightmargin 0pt} 
         \setlength{\itemsep}{0pt} \settowidth 
        {\labelwidth}{#1.}\sloppy}}{\end{list}} 
 
\newcounter{itemlistc} 
\newcounter{romanlistc} 
\newcounter{alphlistc} 
\newcounter{arabiclistc}

\newcommand{\fcaption}[1]{ 
        \refstepcounter{figure} 
        \setbox\@tempboxa = \hbox{\footnotesize Fig.~\thefigure. #1} 
        \ifdim \wd\@tempboxa > 5in 
           {\begin{center} 
        \parbox{5in}{\footnotesize\smalllineskip Fig.~\thefigure. #1} 
            \end{center}} 
        \else 
             {\begin{center} 
             {\footnotesize Fig.~\thefigure. #1} 
              \end{center}} 
        \fi} 
 
\newcommand{\tcaption}[1]{ 
        \refstepcounter{table} 
        \setbox\@tempboxa = \hbox{\footnotesize Table~\thetable. #1} 
        \ifdim \wd\@tempboxa > 5in 
           {\begin{center} 
        \parbox{5in}{\footnotesize\smalllineskip Table~\thetable. #1} 
            \end{center}} 
        \else 
             {\begin{center} 
             {\footnotesize Table~\thetable. #1} 
              \end{center}} 
        \fi} 
 
\def\@citex[#1]#2{\if@filesw\immediate\write\@auxout 
        {\string\citation{#2}}\fi 
\def\@citea{}\@cite{\@for\@citeb:=#2\do 
        {\@citea\def\@citea{,}\@ifundefined 
        {b@\@citeb}{{\bf ?}\@warning 
        {Citation `\@citeb' on page \thepage \space undefined}} 
        {\csname b@\@citeb\endcsname}}}{#1}} 
 
\newif\if@cghi 
\def\cite{\@cghitrue\@ifnextchar [{\@tempswatrue 
        \@citex}{\@tempswafalse\@citex[]}} 
\def\citelow{\@cghifalse\@ifnextchar [{\@tempswatrue 
        \@citex}{\@tempswafalse\@citex[]}} 
\def\@cite#1#2{{$\null^{#1}$\if@tempswa\typeout 
        {IJCGA warning: optional citation argument  
        ignored: `#2'} \fi}}

\def\pmb#1{\setbox0=\hbox{#1} 
        \kern-.025em\copy0\kern-\wd0 
        \kern.05em\copy0\kern-\wd0 
        \kern-.025em\raise.0433em\box0}

\def\fnt#1#2{\footnotetext{\kern-.3em 
        {$^{\mbox{\scriptsize #1}}$}{#2}}} 
 
\def\fpage#1{\begingroup 
\voffset=.3in 
\thispagestyle{empty}\begin{table}[b]\centerline{\footnotesize #1} 
        \end{table}\endgroup} 
 
\def\runninghead#1#2{\pagestyle{myheadings} 
\markboth{{\protect\footnotesize\it{\quad #1}}\hfill} 
{\hfill{\protect\footnotesize\it{#2\quad}}}} 
\font\tenrm=cmr10 
\font\tenit=cmti10  
\font\tenbf=cmbx10 
\font\bfit=cmbxti10 at 10pt 
\font\ninerm=cmr9

\font\eightrm=cmr8

\def\qed{\hbox{${\vcenter{\vbox{                        
   \hrule height 0.4pt\hbox{\vrule width 0.4pt height 6pt 
   \kern5pt\vrule width 0.4pt}\hrule height 0.4pt}}}$}} 
 
\renewcommand{\thefootnote}{\fnsymbol{footnote}}        
 
\begin{document} 
 
\runninghead{Anomalous Higgs Couplings 
$\ldots$} {Anomalous Higgs Couplings  
$\ldots$} 
\normalsize\textlineskip 
\thispagestyle{empty} 
\setcounter{page}{1}

\copyrightheading{}                     
 
\vspace*{0.88truein} 
 
\fpage{1} 
\centerline{\bf ANOMALOUS HIGGS COUPLINGS}  
\vspace*{0.035truein} 
\vspace*{0.37truein} 
\centerline{\footnotesize M.C. GONZALEZ-GARCIA
\footnote{concha.gonzalez@ific.uv.es}} 
\vspace*{0.015truein} 
\centerline{\footnotesize \it Instituto de F\'{\i}sica Corpuscular IFIC/CSIC,}
\baselineskip=10pt 
\centerline{\footnotesize \it Departament de F\'{\i}sica Te\`orica,}
\baselineskip=10pt 
\centerline{\footnotesize\it 
     Universitat de Val\`encia, 46100 Burjassot, Val\`encia, Spain.  } 
\vspace*{0.225truein} 
\publisher{(received date)}{(revised date)} 
\vspace*{0.21truein} 
\abstracts{We review the effects of new effective interactions on the 
Higgs boson phenomenology.  New physics in the electroweak bosonic sector is
expected to induce additional interactions between the Higgs doublet
field and the electroweak gauge bosons leading to anomalous Higgs
couplings as well as to anomalous gauge--boson self--interactions.
Using a linearly realized $SU(2)_L
\times U(1)_Y$ invariant effective Lagrangian to describe  the
bosonic sector of the Standard Model, we review the effects of the new
effective interactions on the Higgs boson production rates and decay
modes.  We summarize the results from searches for the new Higgs
signatures induced by the anomalous interactions in order to constrain
the scale of new physics in particular at CERN LEP and Fermilab Te
vatron colliders.}{}{} 

\textheight=7.8truein 
\setcounter{footnote}{0} 
\renewcommand{\thefootnote}{\alph{footnote}} 
\vspace*{1pt}\textlineskip      
\section{Introduction}  
\noindent 
The Standard Model (SM) of the electroweak interactions based on the
gauge group $SU(2)_L\times U(1)_Y$ has proven to be astonishingly
successful in describing all the available precision experimental
data~\cite{ichep98}. This applies particularly to the predictions of
the couplings of the gauge bosons to the matter fermions. The recent
measurements at LEPII and the Tevatron collider of the gauge--boson
self--couplings, also shed some light on the correctness of the SM
predictions for the interactions on the gauge sector of the theory.

On the other hand, despite we know that the weak gauge bosons, $Z$ and
$W$, are massive and, in consequence, the electroweak gauge symmetry
must be broken at low scales, the precise mechanism of the electroweak
symmetry breaking still remains one of the most important open
questions of the theory. In the SM, the breaking is realized via the
so--called Higgs mechanism in which an scalar $SU(2)$--doublet, the
Higgs boson, is introduced {\it ad hoc} and the symmetry is
spontaneously broken by the vacuum expectation value of the Higgs
field (VEV). In this particular realization, the precise form of the
Higgs couplings to the gauge bosons and the self--couplings of the
scalar are completely determined in terms of one free parameter which
can be chosen to be the Higgs mass $m_{H}$ (or its quartic
self--coupling $\lambda$).  However, in this simple realization, the
theory presents ``naturality'' problems since the running Higgs mass
is quadratically divergent with the scale. This implies the necessity
of large fine--tuning in order to keep the theory perturbative, or,
conversely, the existence of a cut--off scale $\Lambda$ above which
new physics must appear.

Although we do not know the specific form of this theory which will
supersede the SM, we can always parametrize its low--energy effects by
means of an effective Lagrangian~\cite{effective}.  The effective
Lagrangian approach is a model--independent way to describe new
physics that is expected to manifest itself directly at an energy
scale $\Lambda$, larger than the scale at which the experiments are
performed.  The effective Lagrangian depends on the particle content
at low energies, as well as on the symmetries of the low--energy
theory.  For instance, if the electroweak symmetry breaking is due to
a heavy (strongly interacting) Higgs boson, which can be effectively
removed from the physical low-energy spectrum, or to no fundamental
Higgs scalar at all, one is led to consider the most general effective
Lagrangian which employs a nonlinear representation of the
spontaneously broken $SU(2)_L \otimes U(1)_Y$ gauge
symmetry~\cite{Appelquist}. In this case the resulting chiral
Lagrangian is a non-renormalizable non-linear $\sigma$ model coupled
in a gauge-invariant way to the Yang-Mills theory.  If, on the other
hand, there is a light scalar Higgs doublet in the spectrum, the
$SU(2)_L \otimes U(1)_Y$ symmetry must be realized linearly in the
effective theory~\cite{linear}.

In constructing the Lagrangian we will maintain most of the features
of the SM. We will assume that the $\gamma$, $W$ and $Z$ are the gauge
bosons of a $SU(2)_L\times U(1)_Y$ local symmetry which is broken
spontaneously because some order parameter, transforming as a doublet
under $SU(2)_L$, acquires a VEV. In other words, we will consider the
possibility of having a light Higgs boson. Hence, we will use a
linear~\cite{linear,hisz} realization of the $SU(2)_L \times U(1)_Y$
gauge symmetry which for the symmetry breaking sector we parametrized
by a SM--like Higgs doublet field.  In the usual effective Lagrangian
language, at low energy we describe the effects of the new physics,
which will manifest itself directly only at scales above $\Lambda$, by
including in the Lagrangian higher--dimension operators.  In
constructing these operators we will use as building blocks the
gauge--boson and the Higgs fields while keeping the fermionic sector
unchanged. The lowest order operators which can be built without
fermions are of dimension six.

In this review we will concentrate on the effects of the new effective
interactions on the Higgs boson phenomenology and ``vice-versa'' how
the results from searches of the characteristic new signatures for the
Higgs boson induced by the new interactions, can be used to constrain
the scale of new physics.  Anomalous Higgs boson couplings have been
studied in Higgs and $Z^0$ boson decays~\cite{hsz}, and in $e^+
e^-$~\cite{ee,our,our:lep2aaa,our:NLCWWA,our:NLCZZA}, $p
\bar{p}$~\cite{our:tevatronjj,our:tevatronmis,our:tevatron3a} and
$\gamma\gamma$ colliders~\cite{gamma}.

The outline of the paper is as follows. In Sec.~2 we describe the
effective Lagrangian which we will be using and which contains eleven
dimension--six operators with unknown coefficients.  Four of these
operators, ${\cal O}_{\Phi,1}$, ${\cal O}_{DW}$, ${\cal O}_{DB}$, and
${\cal O}_{BW}$, modify the gauge--boson two--point functions at tree
level while three operators, ${\cal O}_{WWW}$, ${\cal O}_{W}$, and
${\cal O}_{B}$, enter at lower order in the gauge--boson three--point
functions.  These operators can be directly constrained by their
tree--level contributions to existing low--energy observables as well
as to the direct gauge--boson production at the Tevatron Collider and
LEPII.  In Secs.~2.2 and 2.3 we summarize these constraints.  Four
operators, ${\cal O}_{WW}$,${\cal O}_{BB}$, ${\cal O}_{W}$, and ${\cal
O}_{B}$, modify the Higgs couplings to the gauge bosons and their
effects can be studied by directly searching for the new Higgs
signatures they induce.  Section 3 presents the effects of these four
operators in the expected production rates of the Higgs boson at
colliders (Sec.~3.3) and in its decay modes (Sec.~3.2).  Section 4
contains the results from the study of specific signatures at the
Tevatron and LEPII.  One of the most interesting features associated
with the presence of these operators is the enhancement of the Higgs
decay rate in two photons that makes the Higgs searches particularly
clean at hadron colliders. The aim of this section is to illustrate
how existing data on some final states containing photons such as :
$p\, \bar p \rightarrow j \, j \, \gamma\, \gamma$, $p\, \bar p
\rightarrow \gamma \,\gamma \,+ \,\not \!\! E_T $, $p\, \bar p
\rightarrow \gamma\, \gamma\, \gamma$ and $e^+\, e^- \rightarrow
\gamma\, \gamma\, \gamma$ at the Tevatron and LEPII can be used to
place limits on the values of the coefficients of the
higher--dimension operators, or, in other words, on the scale of new
physics.

A final discussion of our results will be given in Sec.~5.
Finally in the Appendix we list the relevant Feynman rules for the 
three-- and four--particle interactions. 

\section{Effective Lagrangians}  
\noindent
\subsection{Formalism}
\noindent 
We are interested in the effects at low energy arising from new
physics in the electroweak symmetry--breaking sector.  If $\Lambda$ is
the scale above which new physics will manifest itself directly, we
want to describe the residual effects on the interactions between the
light degrees of freedom of the theory (ie those particles with mass
smaller than $\Lambda$) after integrating out the heavy degrees of
freedom. In the effective Lagrangian language, these effects are
introduced by including in the Lagrangian higher--dimension operators
which will be built out of the relevant light degrees of freedom.
Since we are interested in new physics associated with the electroweak
symmetry-breaking sector, one may expect that the operators involving
fermions are suppressed by powers of the fermions masses over the new
physics scale and can be neglected (with the possible exception of
those associated with the top--quark). Hence in our construction we
assume that the fields mostly affected by these residual interactions
are the three gauge bosons and the doublet Higgs field which we assume
to remain present in the light spectrum and consequently allows us to
use a linear~\cite{linear,hisz} realization of the $SU(2)_L \times
U(1)_Y$ gauge symmetry. Furthermore we also consider that the new
physics respects the parity and charge conjugation symmetries so that
operators which violate $C$ or $P$ can also be neglected.

In the linear representation of the $SU(2)_L \times U(1)_Y$
symmetry breaking mechanism, the SM model is the lowest order
approximation while the first corrections which
can be built involving bosons are of dimension six and
can be written as
\begin{equation}
{\cal L}_{\mbox{eff}} = \sum_n \frac{f_n}{\Lambda^2} {\cal O}_n\; ,
\label{l:eff}
\end{equation}
where the operators ${\cal O}_n$ involve vector boson and/or
Higgs boson fields with couplings $f_n$. This effective
Lagrangian describes well the phenomenology of models that are
somehow close to the SM since a light Higgs scalar doublet, which
we denote as $\Phi$,  is
still present at low energies. There are eleven possible operators
${\cal O}_{n}$ that are $P$ and $C$ even~\cite{linear}.
The building blocks of the new operators are the covariant derivative
\begin{equation}
D_\mu=\partial_\mu +\frac{i}{2} g^\prime B_\mu 
+i g \frac{\sigma_a}{2}W^a_\mu\; ,
\end{equation} 
together with the $U(1)_Y$ and $SU(2)_L$ field strength tensors 
$B_{\mu \nu}$ and $W^a_{\mu \nu}$, 
\begin{eqnarray}
&& \hat{B}_{\mu \nu} = i \frac{g'}{2} B_{\mu \nu}\; , \nonumber\\
&& \hat{W}_{\mu \nu} = i \frac{g}{2} \sigma^a W^a_{\mu \nu} \; .
\end{eqnarray}
where $g$ and $g'$ are the $SU(2)_L$ and $U(1)_Y$ coupling constants
respectively. 

Out of the eleven operators, four of them 
\begin{eqnarray}
&&{\cal O}_{\Phi,1} = \left ( D_\mu \Phi \right)^\dagger \Phi^\dagger \Phi
\left ( D^\mu \Phi \right ) \; , \nonumber \\
&&{\cal O}_{BW} =  \Phi^{\dagger} \hat{B}_{\mu \nu} 
\hat{W}^{\mu \nu} \Phi \; , \label{noblind}  \\ 
&& {\cal O}_{DW} = \mbox{Tr}\left (\left[ D_\mu ,\hat{W}_{\nu \rho} \right]
\left[ D^\mu ,\hat{W}^{\nu\rho} \right]\right)\;, \nonumber \\
&&{\cal O}_{DB}=-\frac{{g^\prime}^2}{2} 
\left(\partial_\mu B_{\nu\rho}\right)
\left(\partial^\mu B^{\nu\rho}\right)\; , \nonumber
\end{eqnarray}
affect the gauge--boson two--point functions at tree level when the 
Higgs field $\Phi$ is replaced by its VEV 
\begin{equation}
\Phi\rightarrow\frac{1}{\sqrt{2}}\left(\begin{array}{l}0\\v \end{array}\right)
\; ,
\label{vev}
\end{equation}
as we discuss below. 

Two of the operators modify only the Higgs self--interactions 
\begin{eqnarray}
&&{\cal O}_{\Phi,2} =\frac{1}{2} 
\partial^\mu\left ( \Phi^\dagger \Phi \right)
\partial_\mu\left ( \Phi^\dagger \Phi \right)\; , \nonumber \\
&&{\cal O}_{\Phi,3} =\frac{1}{3} 
\left(\Phi^\dagger \Phi \right)^3 \; ,
\end{eqnarray}
and they lead to a finite renormalization of the Higgs wave function
and the Higgs potential. In this way, ${\cal O}_{\Phi,1}$ and ${\cal
O}_{\Phi,2}$ induce a finite wave function renormalization of the
Higgs field by a constant $Z_H^{1/2}=[1+(f_{\Phi,1}+2
f_{\Phi,2})v^2/2]^{-1/2}$.  This is the only effect of the operator
${\cal O}_{\Phi,2}$ at one loop. Similarly ${\cal O}_{\Phi,3}$ induces
a finite renormalization of the Higgs potential.

The five remaining operators:
\begin{eqnarray}
&& {\cal O}_{WWW}=\mbox{Tr}[\hat{W}_{\mu \nu}\hat{W}^{\nu\rho}\hat{W}_{\rho}^{\mu}] 
\; , \nonumber \\
&& {\cal O}_{WW} = \Phi^{\dagger} \hat{W}_{\mu \nu} 
\hat{W}^{\mu \nu} \Phi  \; , \nonumber \\
&&{\cal O}_{BB} = \Phi^{\dagger} \hat{B}_{\mu \nu} 
\hat{B}^{\mu \nu} \Phi \; ,  
\label{blind}  \\
&&{\cal O}_W  = (D_{\mu} \Phi)^{\dagger} 
\hat{W}^{\mu \nu}  (D_{\nu} \Phi) \; , \nonumber \\
&&{\cal O}_B  =  (D_{\mu} \Phi)^{\dagger} 
\hat{B}^{\mu \nu}  (D_{\nu} \Phi)  \; , \nonumber 
\end{eqnarray}
contribute to the gauge--boson three-- and four--point functions as
well to the Higgs--gauge--boson couplings. In the Appendix we give the
corresponding Feynman rules for these vertices.  In principle it seems
that the operators ${\cal O}_{WW}$ and ${\cal O}_{BB}$ would also
modify the triple gauge--boson couplings when the Higgs field is
replaced by its VEV (\ref{vev}). However the resulting operators are
proportional to the kinetic energy of the $SU(2)_L$ and $U(1)_Y$ gauge
bosons respectively and therefore they only lead to a finite
renormalization of the gauge fields by constants
$Z_{2B}^{1/2}=[1-f_{BB}v^2/2 ]^{-1/2}$ and
$Z_{2W}^{1/2}=[1-f_{WW}v^2/2 ]^{-1/2}$.
 
\subsection{Low--energy and LEPI constraints}
\noindent 
Some of the operators introduced in the previous section contribute to
low--energy observables and their strength can be constrained by
precision electroweak measurements. They can affect those measurements
through their contributions to both universal~\cite{hisz,hv,hms} (also
called oblique) and non-universal~\cite{our:epsb,other:epsb} (vertex)
corrections.

Operators in Eq.~(\ref{noblind}) modify the oblique corrections to
precision electroweak measurements via their contributions at tree
level to the transverse components of the gauge--boson
propagators. When replacing the Higgs field by its VEV, they lead to
the following bilinear gauge boson interactions
\begin{eqnarray}
{\cal L}& =\frac{\displaystyle 1}{\displaystyle 2\Lambda^2}
\Big\{& f_{DW}~ g^2 \vec{W}_{\mu\nu}\partial^2
\vec{W}^{\mu\nu}
+f_{DB}~ {g'}^2 B_{\mu\nu}\partial^2 B^{\mu\nu} +\nonumber \\
&& f_{BW}~ m_Z^2 s c W^3_{\mu\nu} B_{\mu\nu} + f_{\Phi,1} \frac{v^2}{2} 
m_Z^2 Z^\mu Z_\mu \Big\} \; ,
\label{obliq}
\end{eqnarray}
where $s=\sin\theta_W$ and $c=\cos\theta_W$. 
This contribution to the oblique corrections can be parametrized in
terms of seven parameters~\cite{hisz,hms}, the usual $S$, $T$, and $U$
(or $\epsilon_1$, $\epsilon_2$, $\epsilon_3$)~\cite{obli} together
with four running form factors, such as, for instance, the running of
$\alpha_{QED}$. From Eq.~(\ref{obliq}) one can see that ${\cal
O}_{\Phi,1}$ modifies the $Z$ mass but not the $W$ mass what gives a
contribution to the $\rho=\alpha T=\epsilon_1$ parameter.  ${\cal
O}_{BW}$ induces a mixing between $B$ and $W^3$ and contributes to the
$S$ ({\it i.e.} $\epsilon_3$) parameter. ${\cal O}_{DW}$ and ${\cal
O}_{DB}$ contribute to the running charges and they lead for instance
to an anomalous running of $\alpha_{QED}$ and the weak mixing angle.

Combining the information from precision measurements both at the
$Z$--pole as well as at low energy it is possible to constrain
unambiguously the values of the coefficients of the operators
(\ref{noblind}). We present here the results from a recent analysis
(second article in~\cite{hms}).  The exact limits depend on the values
of $m_{H}$ and $m_{\mbox{top}}$ as the SM prediction depends on those
masses. For $m_{\mbox{top}}=175$ GeV and $90$ GeV$\leq m_{H}\leq 800$
GeV they obtain the following 95\% allowed intervals (in units of
TeV$^{-2}$)
\begin{eqnarray}
&& -1.2 \leq f_{DW}/\Lambda^2\leq 0.56\; , \nonumber \\
&& -33.6 \leq f_{DB}/\Lambda^2\leq 5.6\; , \nonumber \\
&& -1. \leq f_{BW}/\Lambda^2\leq 8.6\; , \label{noblindlim}\\
&& -0.07 \leq f_{\Phi,1}/\Lambda^2\leq 0.61\; . \nonumber 
\end{eqnarray}
These constraints are severe enough to make it very difficult to observe the
effect of these operators in the high--energy observables discussed later.
In what follows we will neglect their effect in our studies.

The five operators in Eq.~(\ref{blind}), also denoted as ``blind'', do
not give tree--level contributions to the low--energy electroweak
precision data. They enter however via one--loop contributions, which,
although generally suppressed by a factor of $1/16\pi^2$ relative to
tree--level effects, can still be large enough to lead to measurable
effects which allows to impose bounds on the corresponding
coefficients. These bounds, however, are far from unambiguous.  One
must bear in mind that without a particular model it is impossible to
predict the interference between the tree--level and the loop--level
corrections as well as possible cancellation among the different
one--loop contributions. The limits presented here (from the analysis
in Ref.~\cite{hms}) are obtained under the ``naturalness'' assumption
that large such cancellation do not occur. In this way, considering
only the effect of one operator at a time, one has the following
constraints at 95\% CL (in units of TeV$^{-2}$)
\begin{eqnarray}
&& -15\leq f_{WWW}/\Lambda^2\leq 25\; , \nonumber\\
&& -12.\leq f_{W}/\Lambda^2\leq 2.5\; ,  \nonumber\\
&& -7.6\leq f_{B}/\Lambda^2\leq 22\; ,  \label{blindlim}\\
&& -24\leq f_{WW}/\Lambda^2\leq 14\; ,  \nonumber\\
&& -79\leq f_{BB}/\Lambda^2\leq 47\; .  \nonumber
\end{eqnarray}
These limits depend in a complicated way on the Higgs mass. The values quoted
above are valid for $m_{H}=200$ GeV.

\subsection{Constraints from gauge--boson production}
\noindent 
Some of the operators introduced in Sec.~2.1 contribute the triple
gauge--boson self--couplings and they can be constrained by the direct
measurements of vector boson pair production processes that have been
conducted by the CDF and D\ \ Collaborations~\cite{wwv:teva} at the
Fermilab Tevatron and by the four LEP~\cite{wwv:lep} experiments at
CERN.

The general form of the  $VW^+W^-$ vertices ($V=Z,\gamma$) in the
presence of the effective Lagrangian (\ref{l:eff}) can be found 
in the Appendix. It involves the coefficients $f_W$, $f_B$, $f_{WWW}$
and $f_{DW}$. When neglecting the contribution from ``not-blind'' operators
($f_{DW}=0$) the triple gauge--boson effective interaction can be rewritten 
as the can be rewritten as the standard 
parametrization~\cite{classical}: 
\begin{eqnarray}
{\cal L}_{WWV} & = &-i g_{WWV} \Bigg\{ 
g_1^V \Big( W^+_{\mu\nu} W^{- \, \mu} V^{\nu} 
  - W^+_{\mu} V_{\nu} W^{- \, \mu\nu} \Big) \nonumber\\ 
& + & \kappa_V W_\mu^+ W_\nu^- V^{\mu\nu}
+ \frac{\lambda_V}{m_W^2} W^+_{\mu\nu} W^{- \, \nu\rho} V_\rho^{\; \mu}
 \Bigg\}
\;,
\label{WWV}
\end{eqnarray}
where $g_{WW\gamma} =  e$,  $g_{WWZ} = e/(s\,c)$. In general these vertices
involve six dimensionless couplings
$g_{1}^{V}$, $\kappa_V$, and $\lambda_V$ $(V = \gamma$ or $Z)$,
after imposing {\it C} and {\it P} invariance.
Electromagnetic gauge invariance requires that $g_{1}^{\gamma} = 1$, while
the other five couplings are related to the new operators according to:
\begin{eqnarray}
\Delta g_1^Z& = g_1^Z-1= &\frac{1}{2}\frac{m_Z^2}{\Lambda^2}f_W \;, 
\nonumber \\
\Delta \kappa_\gamma & = \kappa_\gamma -1 = 
& 1 + \frac{1}{2}\frac{m_W^2}{\Lambda^2}
\Big(f_W + f_B\Big) \;, \label{wwv}\\
\Delta \kappa_Z & = \kappa_Z-1 = & 1 + \frac{1}{2}\frac{m_Z^2}{\Lambda^2}
  \Big(c^2 f_W - c^2 f_B\Big)\;, 
\nonumber \\
\lambda_\gamma &= \lambda_Z = & 
 \frac{3 g^2 m_W^2}{2\Lambda^2}f_{WWW} \;.
\end{eqnarray}
In this case only three of the five couplings remain independent~\cite{hisz}
which can be chosen to be  
$\Delta\kappa_{\gamma}$, $\lambda_{\gamma}$, and $\Delta g_1^Z$. 
The remaining $WWZ$ coupling parameters $\lambda_Z$ and $\Delta\kappa_Z$
are determined by the relations~\cite{hisz}  
\begin{eqnarray}
\lambda_Z = \lambda_{\gamma} & & 
\Delta\kappa_Z = -\Delta\kappa_{\gamma}\tan^{2}\theta_{W} + \Delta g_1^Z\; .
\label{hisz} 
\end{eqnarray} 

A different set of parameters has also been used by the LEP
Collaborations~\cite{lepconv} in terms of 
three independent couplings, $\alpha_{B \Phi}$, $\alpha_{W\Phi}$, 
and $\alpha_{W}$ which simply correspond to the coefficients of
the ${\cal O}_B$, ${\cal O}_W$, and ${\cal O}_{WWW}$ operators 
but defined with a different normalization than the $f_i$ coefficients.
With that normalization 
these parameters are related to the parametrization (\ref{WWV}) 
through 
$\Delta \kappa_\gamma =\alpha_{B \Phi}+\alpha_{W \Phi}$, 
$\Delta g^Z_1= \alpha_{W \Phi}/c_W^2$, and 
$\alpha_{W}=\lambda_\gamma$ while $\lambda_Z$ and $\Delta\kappa_Z$
are determined by the relations (\ref{hisz}).

LEP experiments are sensitive to anomalous triple gauge coupling through the 
$W$--pair cross section, the angular distribution of the produced $W's$ 
and their helicity components which are deduced from the  angles of the
$W$ decay products. In addition single $W$ and single $\gamma$ production 
are also sensitive to the $WW\gamma$ vertex. 

Triple gauge--boson couplings measurements at D\O ~are based on the
analysis of di--boson production events. They obtain limits on
$WW\gamma$ from a fit to the photon $E_T$ spectrum in $W\gamma$ with
the subsequent decay $W\rightarrow \l\nu$.  Limits on $WWZ$ and
$WW\gamma$ couplings are obtained from a fit to the $E_T$ of the two
charged leptons in $p\bar p \rightarrow W^+W^- X\rightarrow\l \nu
\l'\nu' X$ events and from a fit to the $p_T$ spectrum of the
electron-neutrino system in $p\bar p \rightarrow W^+ W^-$ (or $W^\pm
Z$)$X\rightarrow\l \nu jj$.

The enhancement of the gauge--boson--pair cross section from anomalous
couplings increases with the center--of--mass energy, which in
principle gives larger sensitivity to Tevatron experiments compared
with LEP. However, since the backgrounds are also larger at Tevatron,
the overall sensitivity is similar to that of the LEP experiments.
The published results from D\O \ and the four LEP experiments were
combined to produce the tightest available $WW\gamma$ and $WWZ$
coupling limits~\cite{lepd0}.  The 95\% CL limits presented in this
analysis are
\begin{eqnarray}
&&-0.15\leq\Delta\kappa_{\gamma}\leq 0.41\;\;
\mbox{    for $\lambda_\gamma=0$}\; , \\
&&-0.16\leq\lambda_{\gamma}\leq 0.10\;   \;
\mbox{   for $\Delta\kappa_\gamma =0$}\; , 
\nonumber 
\end{eqnarray}
which translates into the following bounds on the coefficients of the
higher--dimension operators in (TeV)$^{-2}$:
\begin{eqnarray}
&&-46\leq (f_W+f_B)/\Lambda^2\leq 127\;  \; \mbox{ for $f_{WWW} =0$}\;, 
 \nonumber \\
&& -41 \leq f_{WWW}/\Lambda^2\leq 26 \; \; \mbox{  for $f_W+f_B=0$}\; . 
\label{limitf:wwv}
\end{eqnarray}
Notice that, since neither $f_{WW}$ nor $f_{BB}$ contribute to the triple
gauge--boson vertices, no direct constraint on these couplings can be 
derived from this analysis. Their first contribution is a modification
of the Higgs couplings which leads to the effects we are going to discuss
next. 
  
\section{Higgs Physics}
\noindent
\subsection{Couplings}
\noindent 
Four of the "blind'' operators in Eq.~(\ref{blind}), 
${\cal O}_{WW}$,${\cal O}_{BB}$, ${\cal O}_{W}$, and  ${\cal O}_{B}$,     
also induce anomalous $H\gamma\gamma$, $HZ\gamma$, $HZZ$ and $HWW$ 
couplings, which, in the unitary gauge, are given by
\begin{eqnarray}
{\cal L}_{\mbox{eff}}^{H} &=& 
g_{H \gamma \gamma} H A_{\mu \nu} A^{\mu \nu} + 
g^{(1)}_{H Z \gamma} A_{\mu \nu} Z^{\mu} \partial^{\nu} H \nonumber \\ 
&+& g^{(2)}_{H Z \gamma} H A_{\mu \nu} Z^{\mu \nu}
+ g^{(1)}_{H Z Z} Z_{\mu \nu} Z^{\mu} \partial^{\nu} H \nonumber \\
&+& g^{(2)}_{H Z Z} H Z_{\mu \nu} Z^{\mu \nu} +
g^{(2)}_{H W W} H W^+_{\mu \nu} W_{-}^{\mu \nu} \; \nonumber \\
&+&g^{(1)}_{H W W} \left (W^+_{\mu \nu} W_{-}^{\mu} \partial^{\nu} H 
+h.c.\right)\, ,
\label{H} 
\end{eqnarray}
where $A(Z)_{\mu \nu} = \partial_\mu A(Z)_\nu - \partial_\nu
A(Z)_\mu$. The effective couplings $g_{H \gamma \gamma}$,
$g^{(1,2)}_{H Z \gamma}$, and $g^{(1,2)}_{H Z Z}$  and 
$g^{(1,2)}_{H WW}$ are related to the coefficients of the 
operators appearing in Eq.~(\ref{blind}) through,
\begin{eqnarray}
g_{H \gamma \gamma} &=& - \left( \frac{g m_W}{\Lambda^2} \right)
                       \frac{s^2 (f_{BB} + f_{WW})}{2} \; , 
\nonumber \\
g^{(1)}_{H Z \gamma} &=& \left( \frac{g m_W}{\Lambda^2} \right) 
                     \frac{s (f_W - f_B) }{2 c} \; ,  
\nonumber \\
g^{(2)}_{H Z \gamma} &=& \left( \frac{g m_W}{\Lambda^2} \right) 
                      \frac{s [s^2 f_{BB} - c^2 f_{WW}]}{c}  \; , 
\label{g} \\ 
g^{(1)}_{H Z Z} &=& \left( \frac{g m_W}{\Lambda^2} \right) 
                      \frac{c^2 f_W + s^2 f_B}{2 c^2} \nonumber \; , \\
g^{(2)}_{H Z Z} &=& - \left( \frac{g m_W}{\Lambda^2} \right) 
  \frac{s^4 f_{BB} +c^4 f_{WW} }{2 c^2} \nonumber \; ,\\  
g^{(1)}_{H W W} &=& \left( \frac{g m_W}{\Lambda^2} \right) 
                      \frac{f_{W}}{2} \nonumber \; , \\
g^{(2)}_{H W W} &=& - \left( \frac{g m_W}{\Lambda^2} \right) 
  f_{WW} \nonumber \; .  
\end{eqnarray}
Notice that, unlike in the SM, the $HZZ$ and $HWW$ couplings are not
proportional to each other as a consequence of the presence of the
operators ${\cal O}_{B}$ and ${\cal O}_{BB}$ which involve the
$U(1)_Y$ strength tensor.  In what follows we will discuss the effects
of the new Higgs couplings in Eq.~(\ref{H}) on the expected Higgs
signals and how existing data can be used to further constraint the
values of the coefficients of these anomalous operators.

\subsection{Decay modes}
\noindent
Higgs decays into gauge bosons are affected by the presence of the 
higher--dimension operators and this  will affect its signature in 
colliders. Larger relative effects are expected 
in the decays $H\rightarrow\gamma\gamma$ and $H\rightarrow Z\gamma$
which in the SM occur only at the one--loop level with contributions
from loops with any massive particle in the loop.
However, the existence of the new interactions (\ref{H}) 
can enhance these widths in a significant way as they contribute
at tree level. These decay widths are given by:
\begin{equation}
\Gamma(H\rightarrow\gamma\gamma)=\frac{m_{H}^3}
{4\pi} \left|g_{H\gamma\gamma}
+\frac{\alpha}{8\pi v}I \right|^2\; , 
\label{width:hgg}
\end{equation}
where $v$ is the Higgs VEV (\ref{vev}) and
$I$ is the complex form factor for the SM contribution to the 
width~\cite{h:gg,h:rev} from loops of all particles with spin 
$i$, charge $e_i$ and colour factor $N_{ci}$,
$I={\displaystyle\sum_i}N_{ci} e_i^2 F_i$. 
$F$ is a complex function
which depends on mass and spin of the particle running in the loop. 
Defining $\tau=4 m_i^2/m_{H}^2$
\begin{eqnarray}
F_1&=&2+3 \tau+3\tau(2-\tau)f(\tau)\; ,\nonumber\\   
F_{1/2}&=&-2\tau[1+(1-\tau)f(\tau)]\;,  \label{Igg}\\
F_0&=&\tau(1-\tau f(\tau))\;.\nonumber 
\end{eqnarray} 
where 
\begin{equation}
f(\tau)=
\left\{\begin{array}{ll}\left[\sin^{-1}(\sqrt{1/\tau}\right]^2\;    \;\;
& \mbox{for  $\tau\geq 1$}\; , \\
\frac{1}{4}\left[ \ln\left(\frac{1+\sqrt{1-\tau}}{1-\sqrt{1-\tau}}\right)
-i\pi\right]^2\;   \;\;
& \mbox{for $\tau\leq 1\; .$} \\ 
\end{array} \right.
\end{equation}
The larger contribution to the SM form factor $I$ comes from the $W$ and 
the top--quark loops being the $W$ contribution  larger by at least 
a factor 4 for $m_{H}>100$ GeV. We can see from Eq.~(\ref{width:hgg}) that
the anomalous contribution is of the order of the SM one for 
$g_{H\gamma\gamma}\sim \frac{\alpha}{8 \pi v}$  (
$f/\Lambda^2\sim {\cal O}$(TeV$^{-2}$)).
\begin{equation}
\Gamma(H\rightarrow\gamma Z)=\frac{m_{H}^3}{16\pi}(1-\frac{x_Z}{4})^3
\left|g^{(1)}_{H Z\gamma}+2 g^{(2)}_{H Z\gamma}+\frac{\alpha}{2\sqrt{2}\pi v} 
A\right|^2\; ,
\label{width:hgz}
\end{equation}
where $x_Z=4 m_Z^2/m_{H}^2$ and $A$ is the form factor for the SM
contribution to the width $A=A_F+A_W$ where $A_W$ gives the
contribution from $W$ loops and $A_F$ from fermion loops, which is
dominated by the top--quark loop, and the explicit form can be found
in Eq.~(2.21) of Ref.~\cite{h:rev}. The value of $A_F$ is small
compared to $A_W$ ($|A_W|/|A_F|>10$ for $m_{H}<300$ GeV).

\begin{figure}[htbp]
\begin{center}
\mbox{\epsfig{file=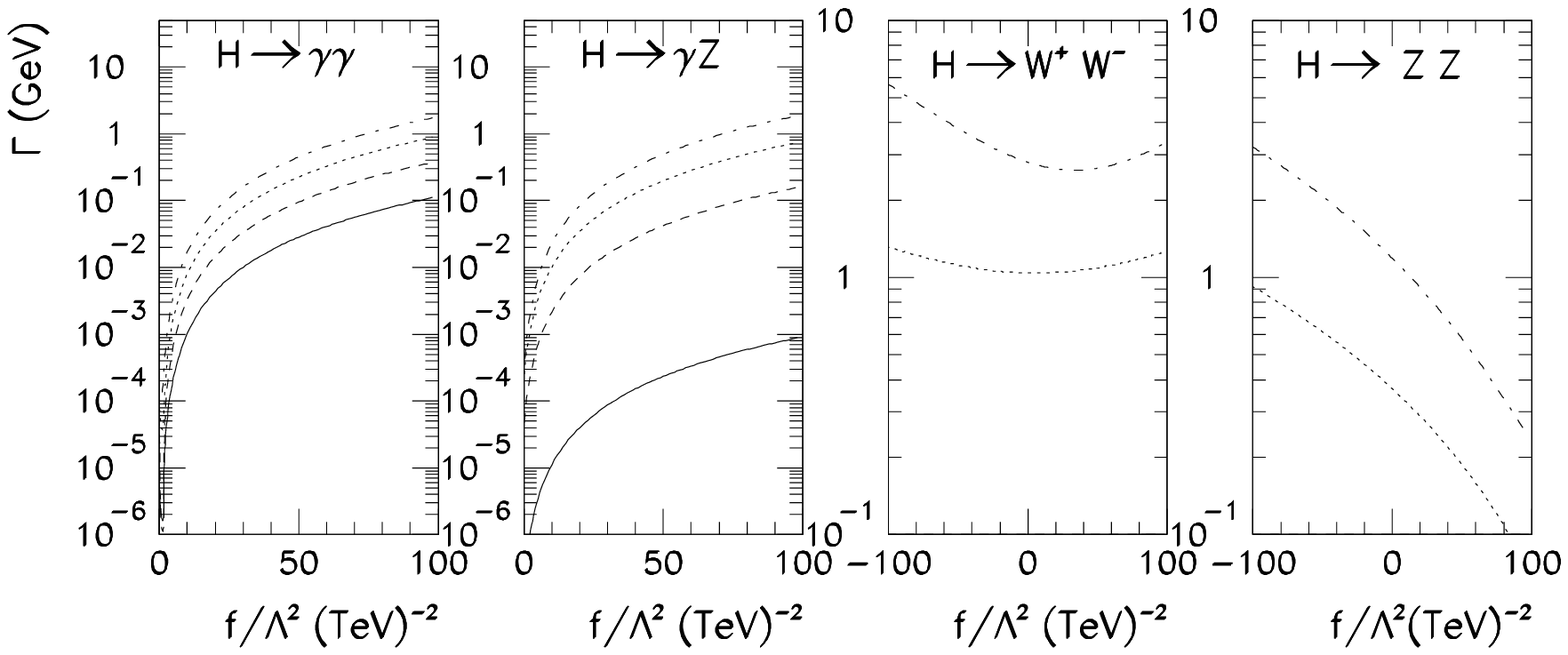,width=\textwidth}}
\end{center} 
\fcaption{Higgs boson decay widths for the different channels as
a function of the anomalous couplings assuming $f_{WW}=f_{BB}=f_W=f_B=f$
while all others are set to zero. The curves correspond to different
Higgs boson mass: $m_{H}=100,150,200,250$ GeV for solid, dashed, 
dotted and dot-dashed respectively.}
\label{fig:widths} 
\end{figure}

\begin{figure}[htbp]
\begin{center}
\mbox{\epsfig{file=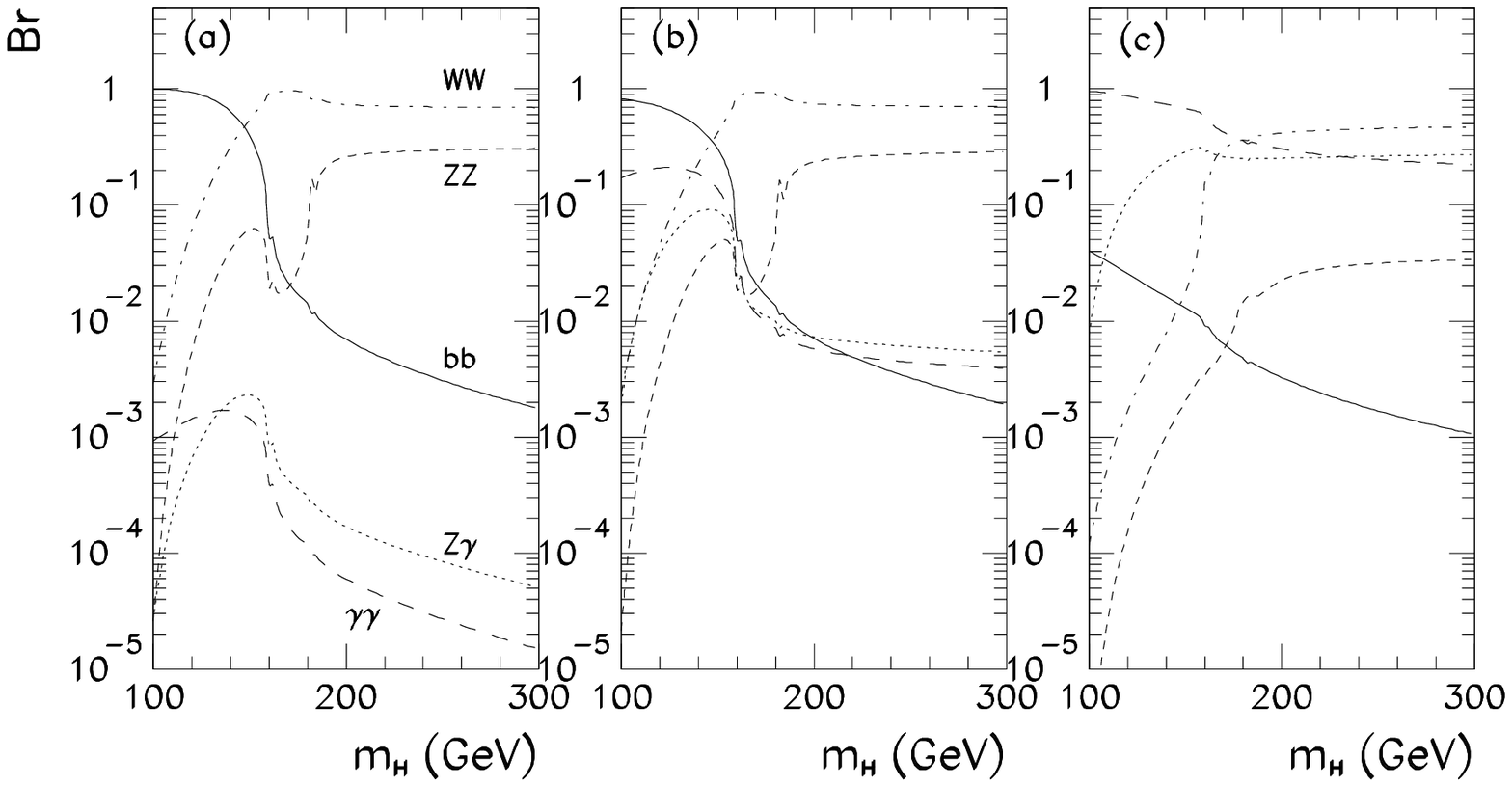,width=\textwidth}}
\end{center} 
\fcaption{Higgs boson decay branching fractions for the different channels as
a function of Higgs mass for different value of the anomalous coupling
$f_{WW}=f_{BB}=f_W=f_B=f$ while all others are set to zero:
{\bf (a)} SM ($f=0$), {\bf (b)} $f/\Lambda^2=10$ TeV$^{-2}$, and 
{\bf (c)} $f/\Lambda^2=100$ TeV$^{-2}$.}
\label{fig:br} 
\end{figure}

In Fig.~\ref{fig:widths} we plot the decay widths for these processes
as a function of the anomalous coefficients and in Fig.~\ref{fig:br}
the corresponding branching ratios as a function of the Higgs mass
for several values of the anomalous coefficients. From the figures
we see that the branching fractions to these decay modes are enhanced 
by 2--4 orders of magnitude as compared to the SM and they can become
dominant for Higgs masses below $W$--pair production threshold. 
Moreover for larger values of $f/\Lambda^2$ they are still a relevant 
decay mode even above  $W$--pair production threshold. 
\noindent 

It is important to notice that the decay mode $H\rightarrow
\gamma\gamma$ only depends on the values of $f_{WW}$ and $f_{BB}$
which are the ``blind'' operators which do not contribute the triple
gauge--boson vertices.  Therefore, it is possible to have significant
enhancements of this decay mode without conflicting with the existing
bounds on anomalous gauge--boson couplings presented in Sec.~2.3.
Conversely, this enhancement, can be used to search for the Higgs
boson in $\gamma\gamma$ signatures, which are cleaner at hadron
colliders. In Sec.~4 we will make use of this enhancement in order to
place bounds on the $f/\Lambda^2$ coefficients from existing data from
the Tevatron and LEPII experiments.

The decay modes into weak--boson pairs, $H\rightarrow ZZ$ and 
$H\rightarrow W^+W^-$ have tree--level contributions from the SM
and therefore the effect of the higher--dimension operators is 
large only for large values of the coefficients. The width for 
these decay modes can be written as:
\begin{equation}
\begin{array}{ll}
{\displaystyle 
\Gamma(H\rightarrow ZZ)=\frac{m_{H}^3}{32\pi}\sqrt{1-x_Z} }
&{\displaystyle 
\Big\{    
 2\left[\frac{x_Z}{2 v}+(x_Z-2) g^{(2)}_{H Z Z}-g^{(1)}_{H Z Z} \right]^2} 
\\[+2.5mm]
&  {\displaystyle 
+ \left[\frac{1}{v}-x_Z(\frac{1}{2 v} +g^{(2)}_{H Z Z})-g^{(1)}_{H Z Z} 
\right]^2\Big\} }\; , 
\end{array}
\label{width:hzz}
\end{equation}
\begin{equation}
\begin{array}{ll}
{\displaystyle 
\Gamma(H\rightarrow W^+W^-)=\frac{m_{H}^3}{16\pi}\sqrt{1-x_W} }
& 
{\displaystyle 
\Big\{    
2\left[\frac{x_W}{2 v}+(\frac{x_W}{2}-1) g^{(2)}_{H W W}
-g^{(1)}_{H WW} \right]^2} \\[+2.5mm]
& 
+{\displaystyle \left[\frac{1}{v}-\frac{x_W}{2}(\frac{1}{v} + g^{(2)}_{H WW})-g^{(1)}_{H WW} 
\right]^2\Big\}}\; .
\end{array} 
\label{width:hww}
\end{equation}
where $x_Z=4 m_Z^2/m_{H}^2$ and $x_W=4 m_W^2/m_{H}^2$.
In these expressions the first term in square brackets corresponds to the
decay into transversely polarized gauge bosons while the second 
corresponds to decay into longitudinally polarized ones. 
We can see from Eq.~(\ref{width:hzz}) that
the anomalous contribution to these widths is of the order of the SM one for 
$g_{HZZ}\sim \frac{1}{v}$ what implies 
$f/\Lambda^2\sim 100$ TeV$^{-2}$.

In Fig.~\ref{fig:widths} we plot the decay widths for these modes as a
function of the anomalous coefficients and in Fig.~\ref{fig:br} the
corresponding branching ratios as a function of the Higgs mass for
several values of the anomalous coefficients.  As expected the effect
in the $W^+W^-$ decay mode is at most a factor two even for very large
values of the $f$ coefficients and the relative effect increases with
$m_{H}$. The effect is a bit larger for the decay $H\rightarrow ZZ$.

\subsection{Production mechanisms at $p\bar p$ and $e^+ e-$ collisions}
\noindent 
Since the Higgs couplings to light fermions, which are the dominant
component of the intial state at $e^+e^-$ and hadron colliders, 
are very weak, the production cross sections for the Higgs boson at colliders 
are in general small. In the SM, production mechanisms with largest 
cross sections are those where the Higgs couples to an intermediate 
heavy particle either a weak gauge bosons or the  top--quark.  
\begin{figure}[htbp]
\begin{center}
\mbox{\epsfig{file=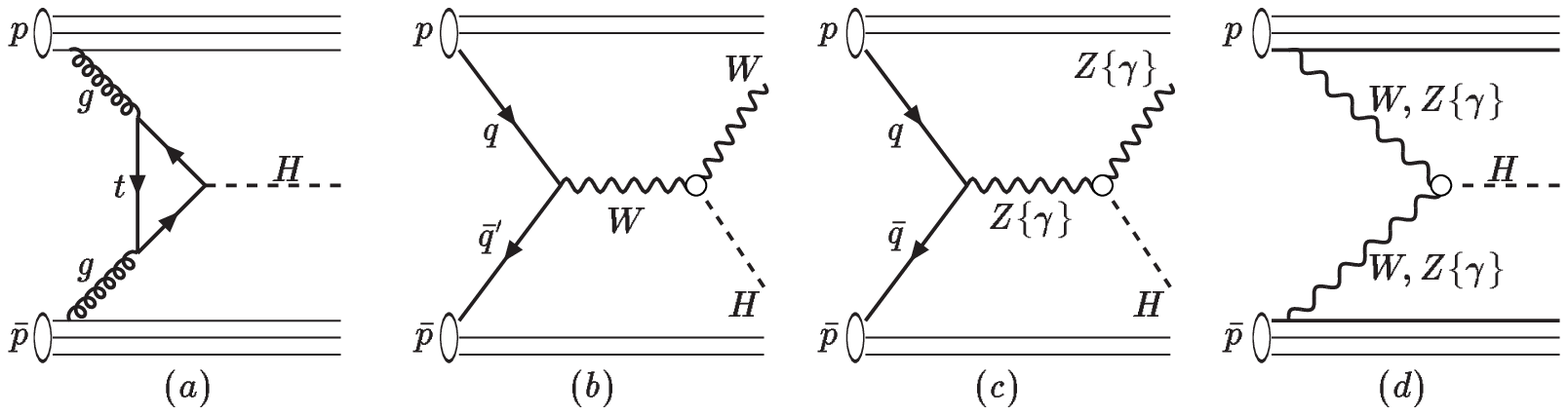,width=\textwidth}}
\end{center} 
\fcaption{Dominant Higgs production mechanisms at hadron colliders: 
{\bf(a)} gluon fusion, {\bf(b)} associated Higgs--$W$ production,
{\bf(c)} associated Higgs--$Z$ production 
(also anomalous Higgs-$\gamma$), and {\bf(d)} gauge--boson
fusion. We denote by a white circle those vertices that are modified
in the presence of the anomalous Higgs couplings (\protect{\ref{H}}) 
and by a particle between brackets those particles whose contributions
only arise for non--vanishing anomalous couplings.}
\label{fig:prodpp} 
\end{figure}

\begin{figure}[htbp]
\begin{center}
\mbox{\epsfig{file=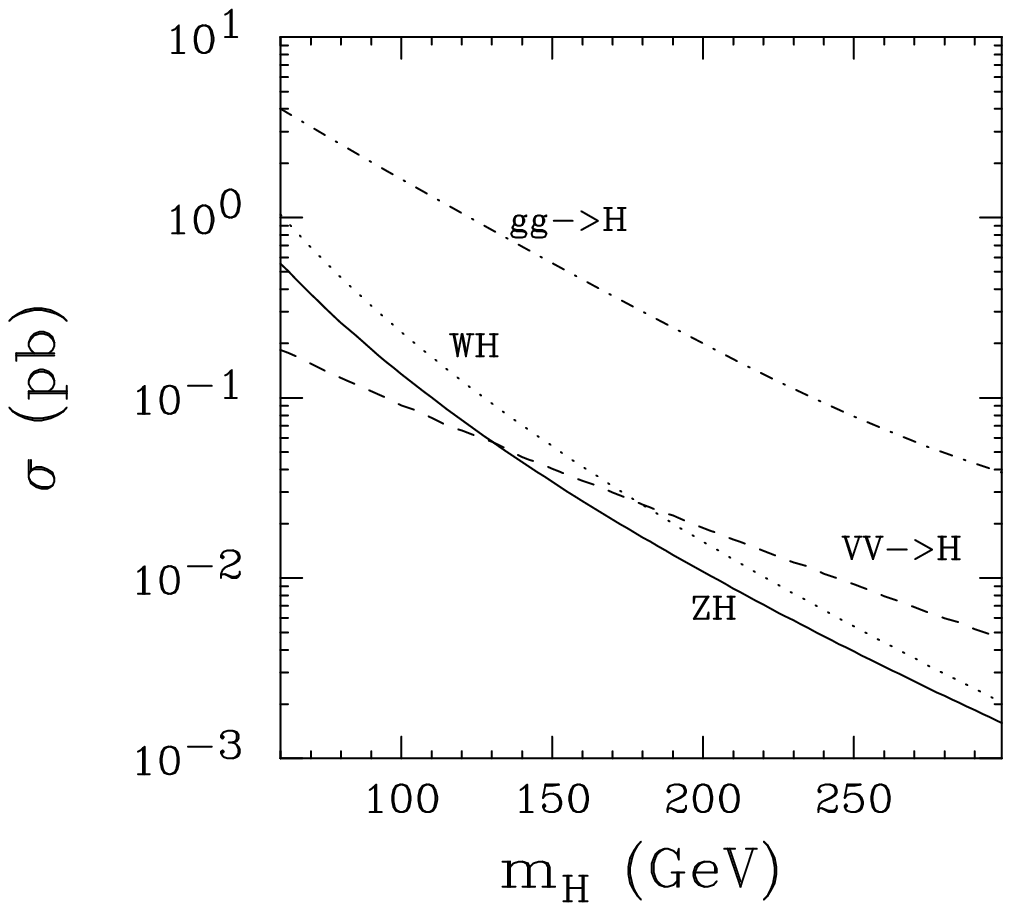,width=0.7\textwidth}}
\end{center} 
\fcaption{Higgs production cross sections at the Fermilab 
Tevatron as a function 
of the Higgs mass in the SM . The curves correspond to different 
production processes as labeled in the figure.}
\label{fig:sigmappsm} 
\end{figure}

In Fig.~\ref{fig:prodpp} we display the dominant production mechanisms for 
the Higgs boson at hadron collisions. For the SM Higgs they are: 
(a) gluon fusion where two initial gluons from the hadrons couple to 
the Higgs boson at one--loop via a virtual top--quark loop, 
(b) associated Higgs production with a $W$ or (c) a  $Z$ boson, and (d) 
$W^+ W^-$ or $ZZ$ fusion. 
In Fig.~\ref{fig:sigmappsm} we plot the cross sections for these processes
at the Tevatron center--of--mass energy of $\sqrt{s}=1.8$ GeV~\cite{h:smw} 
as a function of the Higgs boson mass.  
As seen in Fig.~\ref{fig:sigmappsm}, the production mechanism with 
largest cross section is gluon fusion. However, as we have discussed
before (see Fig.~\ref{fig:br}.a) in the SM the light Higgs boson decays  
dominantly to b--quarks and in consequence the gluon--fusion process  
is swamped by the QCD background of b--quark--pair production since 
there is no other particle in the final state which could be used to 
tag the event. As we have seen in the previous section 
the presence of the new operators modify the Higgs decay modes and 
the Higgs may decay dominantly
into photons. However, still in this case, gluon--fusion production is
swamped by the two--photon background from QED. 

The most important production mechanism for a SM Higgs at the Tevatron
is associated production of the Higgs boson with a $W$ or a $Z$ whose
decay products can be used to trigger the event. On the other hand,
the gauge--boson--fusion processes (d) has similar problems to the
gluon fusion, as it leads to final states with two high--rapidity
jests from the proton remanent plus two b--jets from the Higgs decay
and it is also well below the corresponding QCD background. As we will
see in Sec.~4.2 in the presence of the new operators gauge--boson
fusion can give a significant contribution to the process $p \bar p
\rightarrow jj \gamma\gamma$.

The operators (\ref{blind}) induce new couplings between the Higgs and
the gauge bosons and in consequence modify the SM production processes
cross section by giving rise to new contributions to the SM amplitudes
$q\bar q^{(\prime)}\rightarrow Z^* (W^*)\rightarrow Z (W) H$ as well
as new amplitudes such as $q\bar q\rightarrow \gamma^* \rightarrow Z
H$.  They also provide new production mechanisms such as associated
production of the Higgs boson with a photon (See
Fig.~\ref{fig:prodpp}.c), $q\bar q \rightarrow Z^*,
\gamma^*\rightarrow \gamma H$ which can occur via $g_{H \gamma
\gamma}$ as well as $g^{(1)}_{H Z \gamma}$ and $g^{(2)}_{H Z \gamma}$.
Higgs production by gauge--boson fusion is also modified as now also
$Z\gamma$ and $\gamma\gamma$ fusion are possible (See
Fig.~\ref{fig:prodpp}.d).

\begin{figure}[htbp]
\begin{center}
\mbox{\epsfig{file=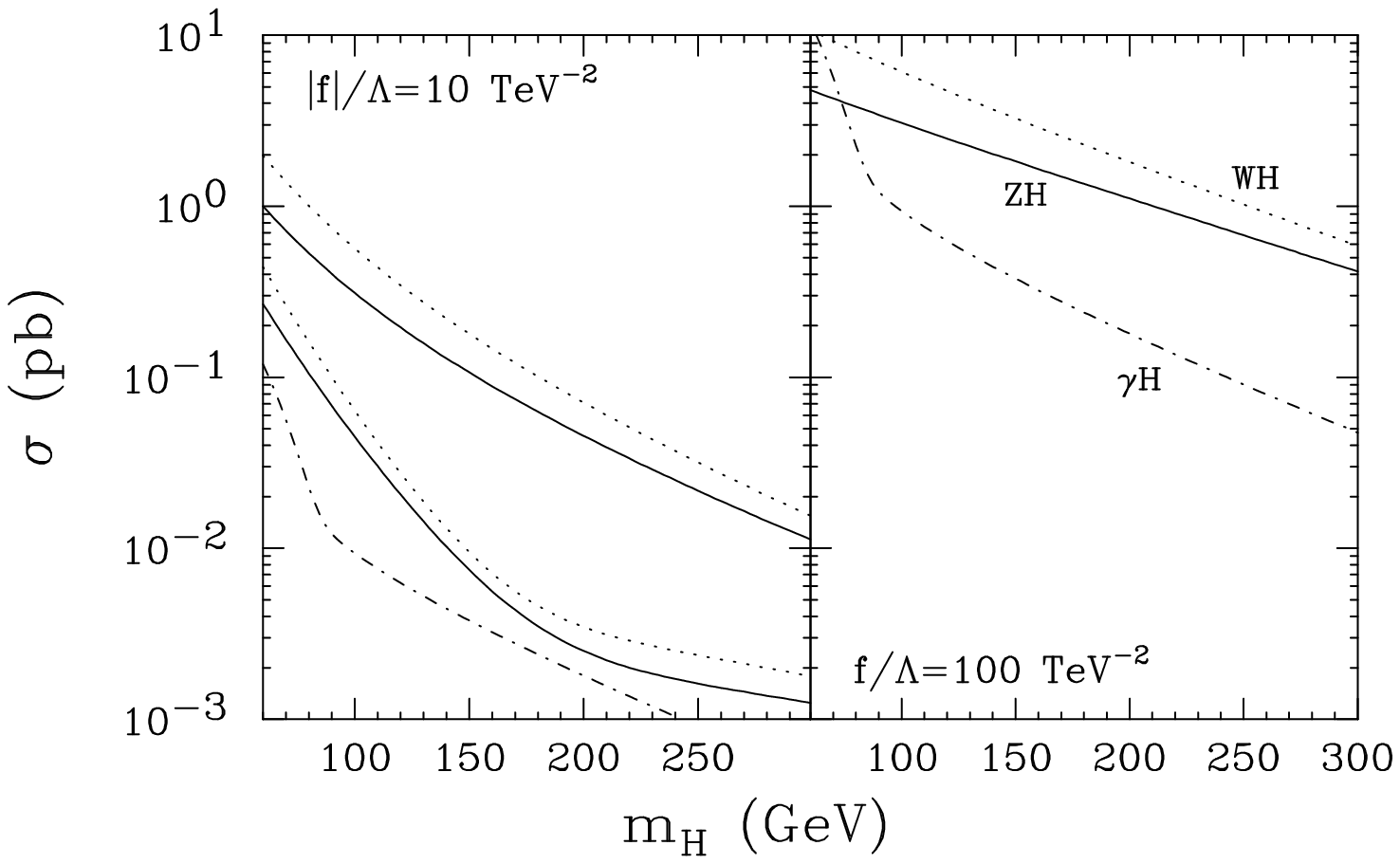,width=0.9\textwidth}}
\end{center} 
\fcaption{Associated Higgs production cross sections at the Fermilab Tevatron 
as a function of the Higgs mass in the presence of anomalous couplings.
The left figure corresponds to $f_{WW}=f_{BB}=f_W=f_B=f$ with $|f|=10$
TeV$^{-2}$.  The two upper curves are for $ZW$ and $ZH$ productions
for $f=-10$ TeV$^{-2}$ while the lower ones are for $f=10$
TeV$^{-2}$. The associated $\gamma H$ cross section (dash-dotted) line
is practically independent on the sign of $f$. The right figure
corresponds to $f=100$ TeV$^{-2}$.}
\label{fig:sigmappano} 
\end{figure}

In Fig.~\ref{fig:sigmappano} we plot the Higgs production 
cross section for the associated production processes in the
presence of the new operators. 
For the purpose of illustration we assume all the four coefficients
equal $f_{WW}=f_{BB}=f_W=f_B=f$.  
As seen in Fig.~\ref{fig:sigmappano}, for large values of the anomalous 
coefficients, the presence of the new operators yields production 
rates larger than in the SM. However 
for intermediate values, the associated Higgs--$Z$ and Higgs--$W$ 
production cross 
sections can be smaller than in the SM as a consequence of the 
possible destructive interference between the SM amplitudes and 
some of the anomalous ones as seen in the left panel in 
Fig.~\ref{fig:sigmappano}. We also see that 
associated Higgs--$\gamma$ production is 
particularly dominant for Higgs bosons lighter than the 
$Z$ when an on-shell $Z$ can decay into $\gamma H$. 

At $e^+ e^-$ collisions the dominant production mechanism
for the SM Higgs are associated Higgs--$Z$ production (also
known as ``Higgs-strahlung''), $e^+e^- \rightarrow ZH$, and $W^+W^-$
fusion $e^+e^-\rightarrow \nu_e\bar\nu_e H$. At the LEPII energies
the gauge--fusion process has a considerably smaller  cross section
as it is suppressed by an additional power of the electroweak coupling
constant.  

\begin{figure}[htbp]
\begin{center}
\mbox{\epsfig{file=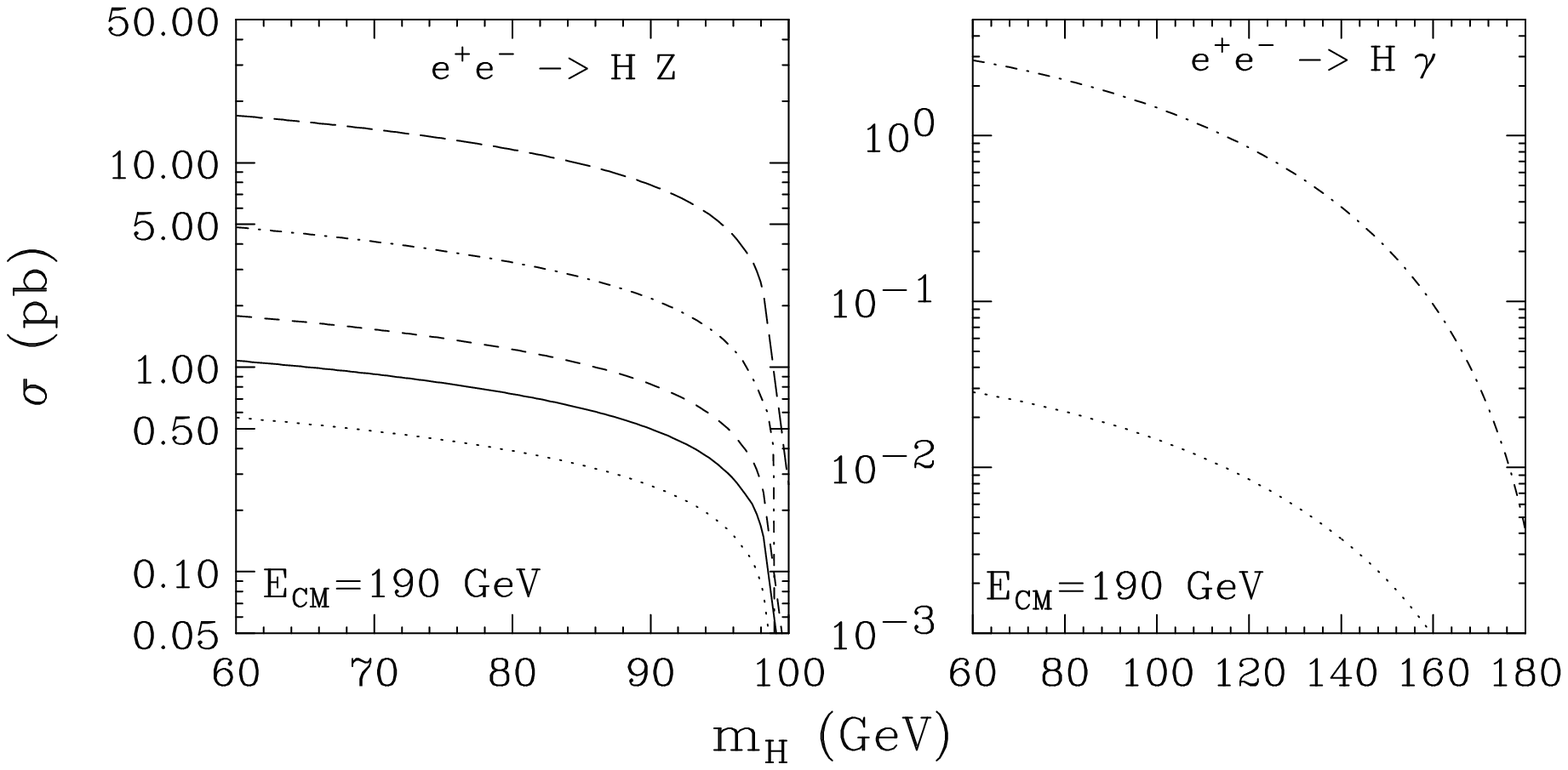,width=\textwidth}}
\end{center} 
\fcaption{Higgs production cross sections at LEPII as a function 
of the Higgs mass. The curves correspond to different values of the 
anomalous couplings assuming $f_{WW}=f_{BB}=f_W=f_B=f=0,10, -10, 100, -100$
TeV$^{-2}$ for solid, dotted dashed, dash-dotted, and long-dashed 
respectively. Notice that for $e^+ e^-\rightarrow \gamma H$ the
cross section is almost independent of the sign of the anomalous
coupling and we are plotting the cross section only for positive values: 
$f/\Lambda^2=10 (100)$ TeV$^{-2}$ for dash (dash-dotted).}
\label{fig:sigmaee} 
\end{figure}

The operators (\ref{blind}) give rise to new contributions to the SM
amplitude for Higgs--$Z$ production as well as new amplitudes: $e^+
e^-\rightarrow Z^* \gamma^* \rightarrow Z H$.  They also provide the
possibility of associated production of the Higgs boson with a photon,
$e^+ e^- \rightarrow Z^*, \gamma^*\rightarrow \gamma H$.  In
Fig.~\ref{fig:sigmaee} we plot the cross sections for these processes
at the center--of--mass energy $\sqrt{s}=190$ GeV.
Again we see that for intermediate values of the anomalous coupling
$f$, the production cross section for Higgs--$Z$ can be smaller than in the
SM as a consequence of the negative interference between the SM and
the anomalous amplitudes. For large values of the coupling, however
the cross section is always larger than in the SM.

\section{Study of Specific Processes}
\noindent
\subsection{Introduction}
\noindent
In this section we are going to analyze some processes which can be
used to study the effect of the dimension--six 
operators in Higgs searches. The aim is to illustrate how by using 
data already collected by the experiments at Tevatron and LEPII
it is possible to obtain further constraints on the values of the
coefficients of the operators (\ref{blind}). In Sec.~4.6
we will show the final results of the combination of all these searches
and compare them with the limits discussed in Sec.~2.3.

We first consider Higgs production at the Fermilab
Tevatron and CERN LEPII collider with its subsequent decay into two photons. 
We explore the signatures:
\begin{eqnarray}
p\, \bar p & \rightarrow & j \, j \, \gamma\, \gamma\; , \nonumber  
\\
p\, \bar p & \rightarrow & \gamma \,\gamma \,+ \,\not \!\! E_T \nonumber\; , \\
p\, \bar p & \rightarrow & \gamma\, \gamma\, \gamma\; , \nonumber\\
e^+\, e^-  & \rightarrow & \gamma\, \gamma\, \gamma\; . \label{proc}
\end{eqnarray} 
This type of events containing two photons plus missing energy, additional
photons or charged fermions represent a signature for several
theories involving physics beyond the SM, such as some classes of
supersymmetric models~\cite{xer} and they have been
extensively searched for~\cite{opal:ggg,d0jj,d0miss,cdf}.
In the framework of anomalous
Higgs couplings presented before, they can also arise from the
production of a Higgs boson which subsequently decays in two
photons. As we have seen in Sec.~3.2 
in the SM, the  decay width $H \to \gamma \gamma$ is very
small since it occurs just  at one--loop level~\cite{h:gg}.
However, the existence of the new interactions (\ref{H}) 
can enhance this width in a significant way.  
Recent analyses of these signatures presented a good agreement with 
the expectations from the SM. Thus we can employ these negative experimental 
results  to constrain new anomalous couplings in the bosonic sector of the
SM. We will also discuss the expected sensitivity at the Tevatron collider
with increased luminosity as well as in the Next Linear Collider (NLC).

In the calculations presented below, we have included all SM (QCD plus
electroweak), and anomalous contributions that lead to these
final states. The SM one--loop contributions to the $H\gamma\gamma$
and $H Z\gamma$ vertices are introduced through the use of the 
effective operators with the corresponding form factors 
(\ref{Igg}) and
(\ref{width:hgz}) in the coupling. Neither the narrow--width approximation 
for the Higgs boson 
contributions, nor the effective--$W$--boson approximation were employed. In 
this way, the effect of all interferences between the
anomalous signature and the SM background has been 
consistently included. The SM Feynman diagrams corresponding to
the background subprocess can be generated by Madgraph~\cite{madgraph} 
in the framework of Helas~\cite{helas}. The
anomalous couplings arising from the Lagrangian
(\ref{l:eff}) were implemented in Fortran routines and were
included accordingly. For calculations of processes at the Tevatron, 
we have used the MRS (G)~\cite{mrs} set of
proton structure functions with the scale $Q^2=\hat{s}$.

In order to compare the theoretical predictions with  the experimental 
results from the different searches described below, one must 
compute the expected number of events 
after including the experimental cuts on the final state particles.
Some of the cuts take into account the geometrical
acceptance of the detector and others are designed to improve
the sensitivity to the signal while reducing
the possible backgrounds. The cuts employed in this review are usually 
defined in terms of the following kinematical variables 
\begin{equation}
\begin{array}{ll}
\mbox{pseudorapidity} & {\displaystyle \eta=-\ln\frac{\theta}{2}} \\[+2mm]
\mbox{trasverse momentum} & p_T=\sqrt{E^2-m^2-p_L^2} \\[+2mm]
\mbox{trasverse mass} & E_T=\sqrt{E^2-p_L^2}\\[+2mm]
\mbox{invariant mass of the system $ab$} &M^2_{ab}=(p_a+p_b)^2 \\[+2mm]
\mbox{separation between $a$ and $b$} & 
\Delta R_{ab}=\sqrt{(\eta_a-\eta_b)^2+(\phi_a-\phi_b)^2 }
\end{array}
\end{equation}
where $p_{a}$ is the four--momentum of particle $a$, $E$ is the particle 
energy, $p_L$ is the projection of the particle three--momentum along the 
beam axis, $\theta$ is polar the scattering angle, $\tan\theta=p_T/p_L$, and 
$\phi$ is the azimuthal angle of the momentum.

The coupling $H\gamma\gamma$ derived in (\ref{g})
involves $f_{WW}$ and $f_{BB}$~\cite{hsz}.  In consequence, the
anomalous signature for any of the processes in Eq.~(\ref{proc}) 
is only possible when
those couplings are not vanishing. The couplings $f_B$ and $f_W$,
on the other hand, affect the production mechanisms for the Higgs
boson. In what follows, we are going to present the results of 
the analysis of the processes (\ref{proc}) for two different
scenarios of the anomalous coefficients: 
\begin{itemize}
\item $(i)$ Suppressed $VVV$ couplings compared to the
$H\gamma\gamma$ vertex: $f_{BB,WW} \gg f_{B,W}$,
\item $(ii)$ All coupling with the same magnitude:
$f_{BB,WW,B,W} = f$.
\end{itemize}
In order to establish the attainable bounds on the coefficients,
an upper limit on the number of signal events based on
Poisson statistics is imposed. In the absence of background this implies
$N_{\mbox{signal}} < 1 \,(3)$ at 64\% (95\%) CL. In the presence
of background events, the modified Poisson 
analysis~\cite{otaviano} was employed. 

\subsection{$p\bar p\rightarrow jj\gamma\gamma$}
\begin{figure}[htbp]
\begin{center}
\mbox{\epsfig{file=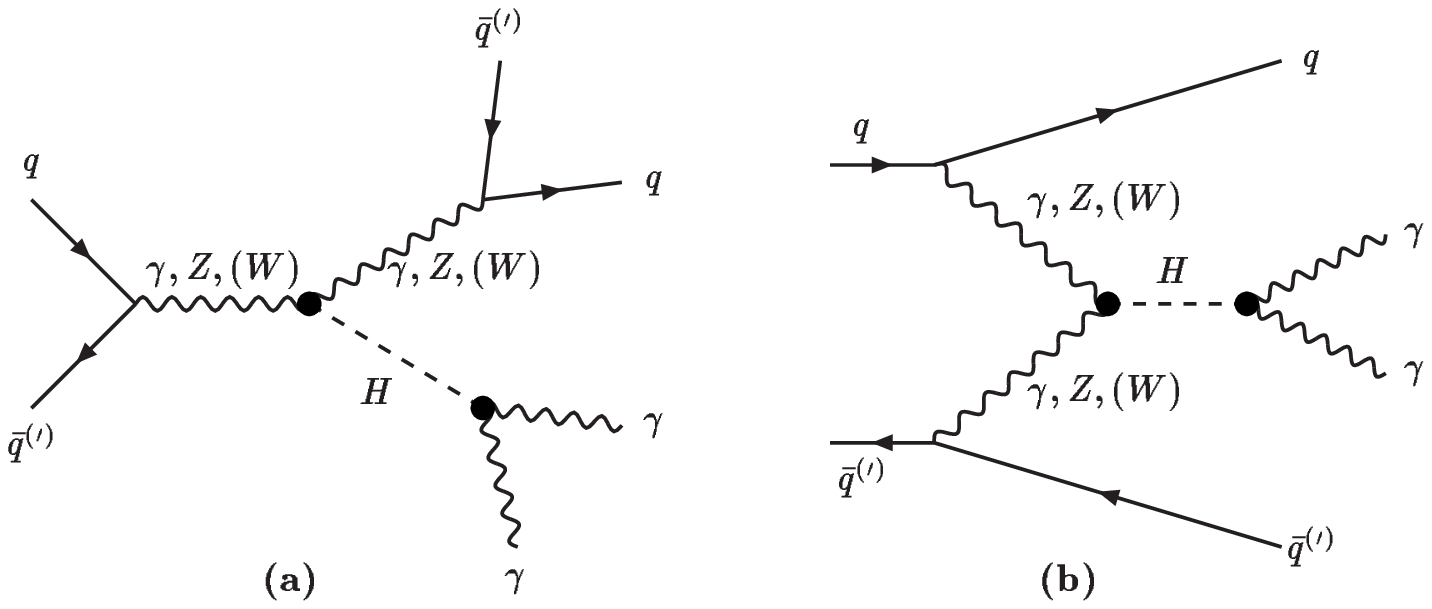,width=\textwidth}}
\end{center} 
\fcaption{Dominant anomalous contribution for the $\gamma\gamma jj$
production: {\bf (a)} Associated production, {\bf(b)} Gauge--boson fusion.} 
\label{fig:jjgg:diag} 
\end{figure}

\noindent
Let us start with the analysis of the process 
$p\bar p\rightarrow j \, j \, \gamma \gamma$. 
A total of 1928 SM amplitudes are involved in the different 
subprocesses contributing to this signature~\cite{jjgg} while
236 anomalous amplitudes~\cite{our:tevatronjj} are generated by 
the operators (\ref{blind}). All these amplitudes are computed
numerically as described above.
The dominant anomalous contribution to this final state when the
two photons have large invariant mass arises from Higgs--$V$ (where V
is any of the  gauge boson) associated 
production and from gauge--boson fusion and are displayed in 
Fig.~\ref{fig:jjgg:diag}.

The D\O ~Collaboration at the Fermilab Tevatron have searched for 
$p \bar p\rightarrow \gamma\gamma j j$ events with high  two--photon
invariant masses~\cite{d0jj} using the 100 pb$^{-1}$ 
of integrated luminosity collected in Run I. They report that 
no event with two--photon invariant mass in the range $60
< M_{\gamma\gamma} < 220$ was observed.
In order to use this D\O ~result one must 
compute the expected number of events in the presence of the anomalous
coefficients after including the same cuts 
on the final state particles which take into account the geometrical
acceptance of the detector and which are also aimed to maximally reduce
the possible backgrounds as discussed below:
\begin{equation}
\begin{array}{ll}
\mbox{For the photons} & \\[0.1cm]
|\eta_{\gamma 1}|< 1.1 \mbox{  or  } 1.5<|\eta_{\gamma 1}|<2 & 
p_T^{\gamma 1}>20 \mbox{ GeV} \\
|\eta_{\gamma 2}|< 1.1 \mbox{  or  } 1.5<|\eta_{\gamma 2}|<2.25 & 
p_T^{\gamma 2}>25 \mbox{ GeV} \\
\sum \vec p_T^\gamma >10 \mbox{ GeV} &  \\[0.1cm]
\mbox{For the jets} & \\ [0.1cm]
|\eta_{j 1}|< 2  & 
p_T^{j 1} > 20 \mbox{ GeV} \\
|\eta_{j 2}|< 2.25 & 
p_T^{j2}> 15 \mbox{ GeV} \\
\sum \vec p_T^j >10 \mbox{ GeV} & R_{\gamma j} > 0.7  \\
40\le M_{jj} \le 150 \mbox{ GeV}
\end{array}
\label{jjgg:cuts}
\end{equation} 
Since we are not working in the narrow width approximation we must
also impose the reconstruction of the photon invariant mass
around the Higgs mass. In order to do so we 
assume an invariant--mass resolution for the
two photons of  $\Delta M_{\gamma\gamma}/ M_{\gamma\gamma} =
0.15/ \sqrt{M_{\gamma\gamma}} \oplus 0.007$~\cite{h:smw}. Both
signal and background were integrated over an invariant--mass bin
of $\pm 2 \Delta M_{\gamma\gamma}$ centered around $m_{H}$.

\begin{figure}[htbp]
\begin{center}
\mbox{\epsfig{file=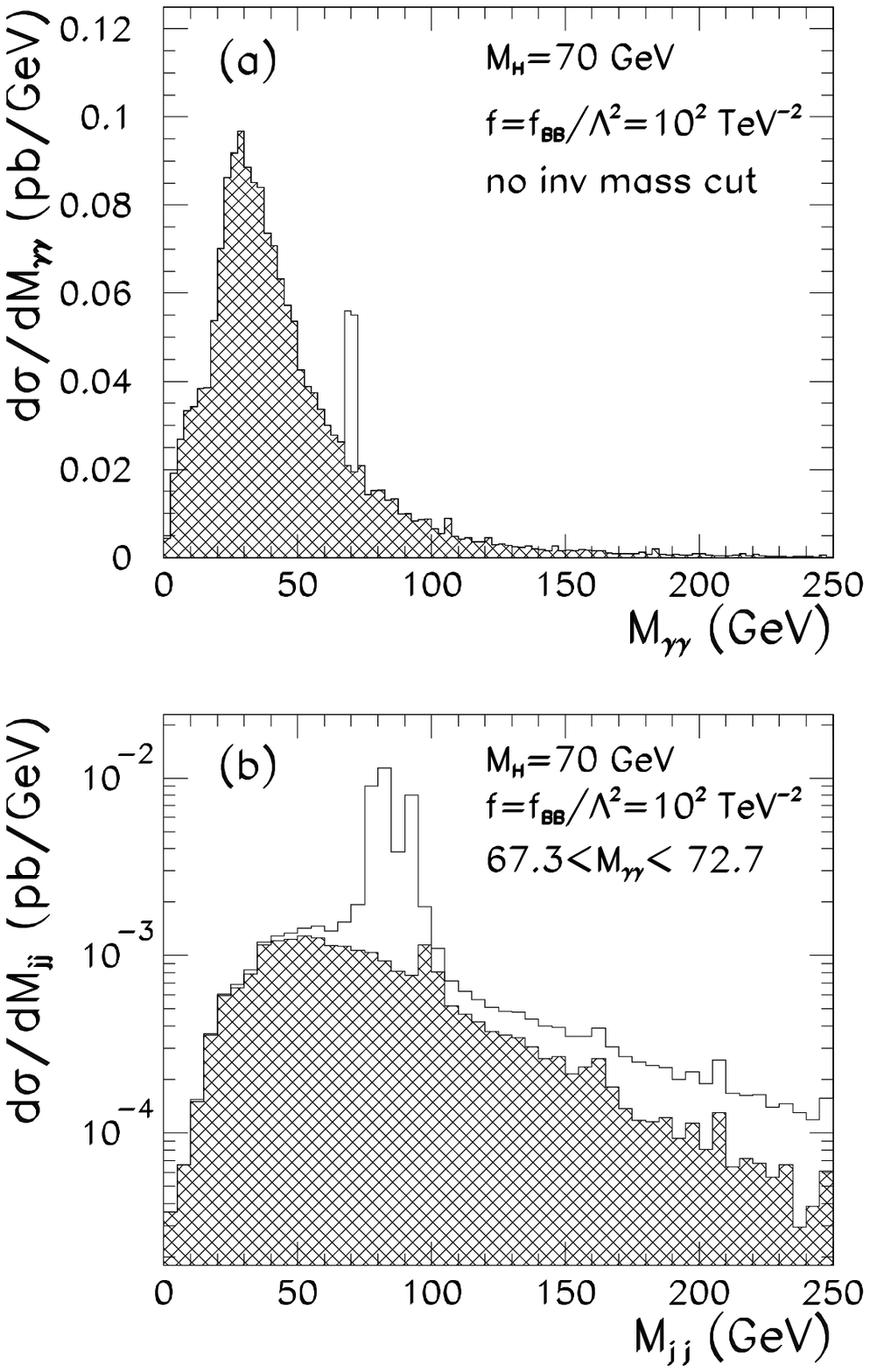,width=0.4\textwidth}}
\end{center} 
\fcaption{(a) Two photon invariant mass distribution for the
background (shaded histogram) and for the signal (clear
histogram) before applying any cut, for $m_{H}=70$ GeV and
$f_{BB}/\Lambda^2 = 100$ TeV$^{-2}$. (b) Two jet invariant mass
distribution, after the cut on the two photon invariant mass.}
\label{fig:jjgg:inv} 
\end{figure}

As mentioned above the anomalous signature receives
contributions from both Higgs associated production and gauge--boson fusion.
For the sake of illustration, we show in Fig.~\ref{fig:jjgg:inv}.a the
invariant mass distribution of the two photons for $m_{H}=70$ GeV
and $f_{BB}/\Lambda^2 = 100$ TeV$^{-2}$, without any cut on
$M_{\gamma\gamma}$ or $M_{jj}$. We see a peak of events due to the on--shell
Higgs production.  
Figure.~\ref{fig:jjgg:inv}.b displays the invariant mass distribution 
of the jet pair 
after imposing the Higgs mass reconstruction on the $\gamma$--$\gamma$ system.
One can clearly see that there is a significant excess of events in 
the region  $M_{jj}
\sim m_{W,Z}$ corresponding to the process of associate
production (Fig.\ref{fig:jjgg:diag}.a). It is also possible to distinguish the
tail  corresponding to the Higgs production from $WW/ZZ$ fusion
(Fig.\ref{fig:jjgg:diag}.b) for $M_{jj} > 100$ GeV.  
One can isolate the majority
of events due to associated production, and the corresponding
background, by integrating over a bin centered on the $W$ or $Z$
mass, which is equivalent to the two--jets--invariant--mass cut listed above. 

After imposing all the cuts, one gets a reduction on the signal
event rate which depends on the Higgs mass. For instance, 
the geometrical acceptance and
background rejection cuts (\ref{jjgg:cuts}) 
account for a reduction factor of 15\% for
$m_{H}=60$ GeV rising to 25\% for 
$m_{H}=160$ GeV. One must also include in
the analysis the particle identification and trigger
efficiencies. For leptons and photons they vary from 40\% to 70\%
per particle~\cite{D0,D02}. For the $jj\gamma\gamma$ final
state we estimate the total effect of these  efficiencies to be
35\%. We therefore obtain an overall efficiency for the
$jj\gamma\gamma$ final state of 5.5\% to 9\% for $m_{H} =
60$--$160$ GeV. 

Next one must consider the possible backgrounds which contribute to
the same final state. The dominant physics background is a mixed 
QCD--QED process which is automatically included since the calculation
is done adding all the SM plus anomalous amplitudes. 
When the cuts (\ref{jjgg:cuts}) and the efficiencies discussed above 
are included, this background is reduced to less than 0.2 events for 
the present luminosity. 
Dominant backgrounds, however, are due to misidentification
when a jet fakes a photon. The probability for a jet to fake a
photon has been estimated to be of a few times $10^{-4}$~\cite{D0}. 
Although this probability is small, it becomes the
main source of  background for the $j j \gamma\gamma$ final state
because of the very large multijet cross section. 
In Ref.~\cite{d0jj} this background is estimated to lead to $3.5\pm
1.3$ events with invariant mass  $M_{\gamma\gamma}>60 $ GeV and
it has been consistently included in the derivation of the
attainable limits presented below. 

The results of this analysis can be used to place limits on the 
coefficients of the higher--dimension operators. Since D\O ~report  
that no event with two--photon invariant mass in the range $60
< M_{\gamma\gamma} < 220$ has been observed, 
a $95\%$ CL in the determination of the anomalous coefficient 
$f_i$, $i=WW, BB, W, B$ is attained requiring 3 events
coming only from the anomalous contributions. 

\begin{figure} [htbp]
\centerline{\epsfig{file=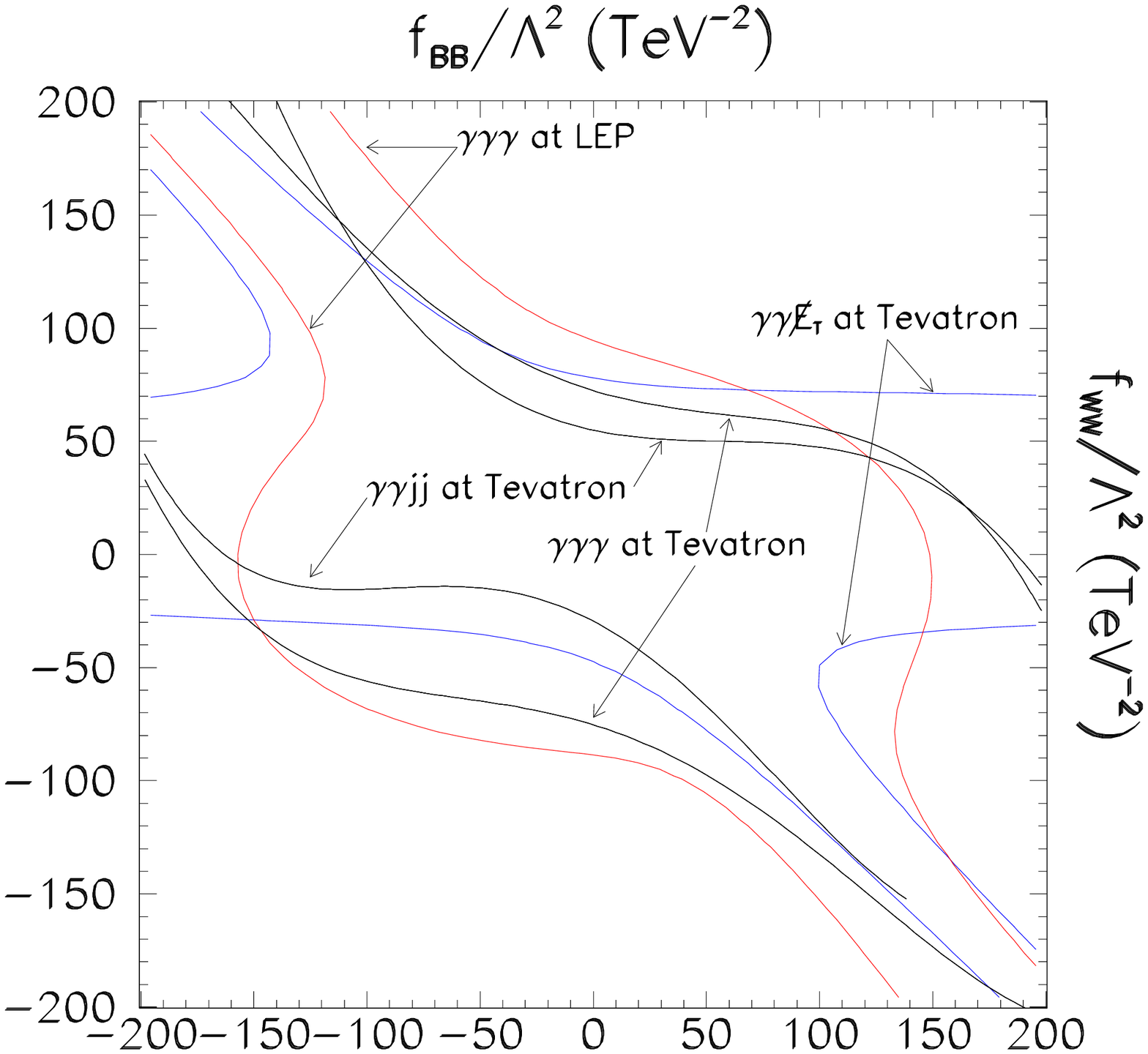,width=0.5\textwidth}
            \epsfig{file=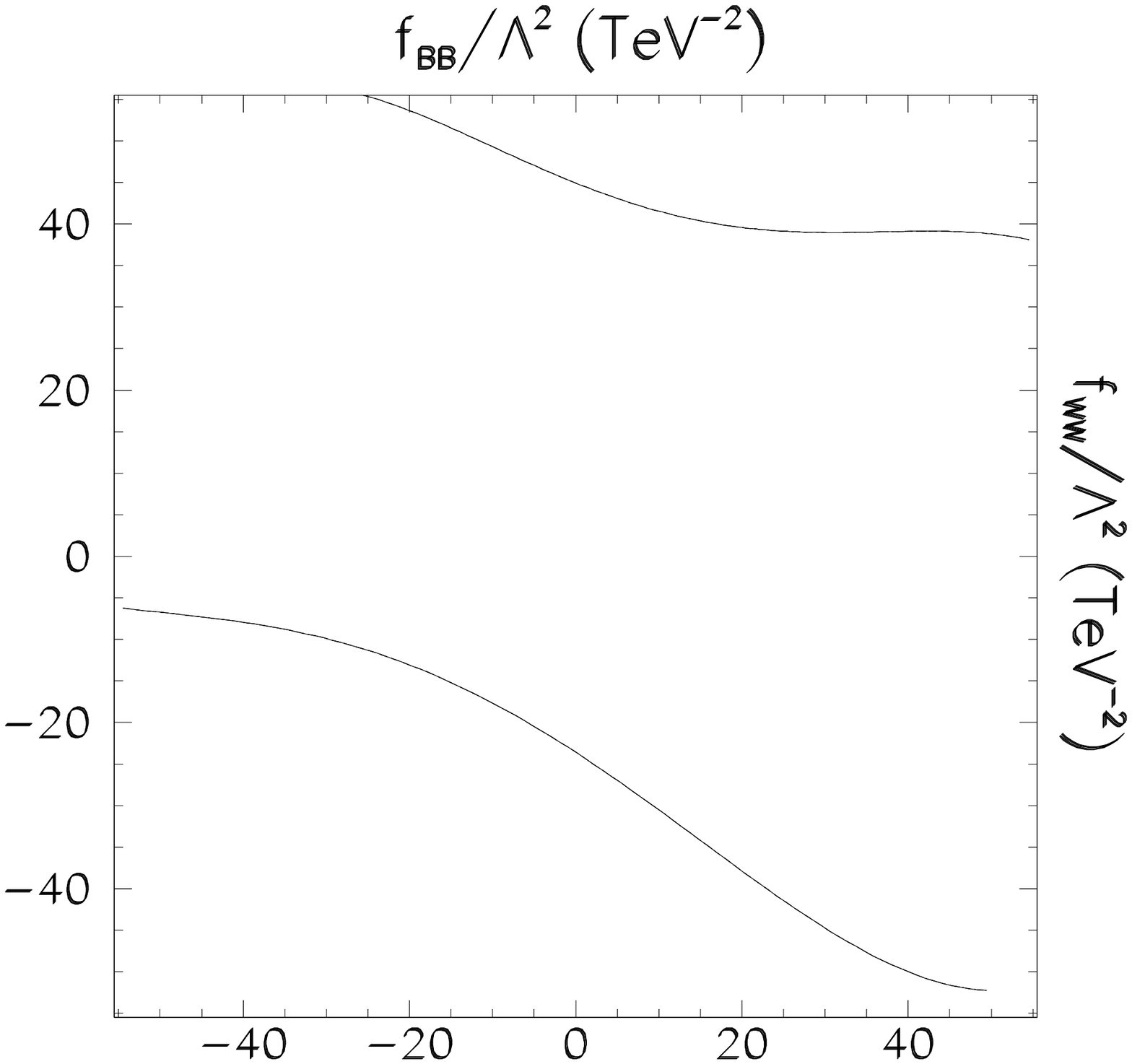,width=0.5\textwidth}}
\fcaption{{\bf (a)} Exclusion region outside the curves in the
$f_{BB} \times f_{WW}$ plane, in TeV$^{-2}$, based  on the D\O
~analysis \protect~\cite{d0jj} of $\gamma\gamma j j$ production,
on  the D\O ~analysis \protect~\cite{d0miss} of $\gamma\gamma \not
\!\! E_T$, on the CDF analysis \protect~\cite{cdf} of
$\gamma\gamma\gamma$ production, and on the OPAL analysis
\protect~\cite{opal:ggg} of $\gamma\gamma\gamma$ production,
always assuming $ m_{H} = 100$ GeV.  The curves show the 95\% CL
deviations from the SM total cross section. {\bf (b)} Same as
{(a)} for the combined analysis. Notice the change of scale
in the last plot.}
\label{fig:fww:fbb}
\end{figure}

In Fig.~\ref{fig:fww:fbb}.a we present the  
region in the $f_{WW}$, $f_{BB}$  plane that can be excluded 
at 95 \% CL in scenario $(i)$, this is, assuming that these
are the only non-vanishing couplings, for $m_{H}=100$ GeV.  
Since the anomalous contribution to 
$H\gamma\gamma$ is zero for
$f_{BB} = - f_{WW}$, the bounds become very weak close to this
axis, as clearly shown in Fig.~\ref{fig:fww:fbb}.  
We should remind that these couplings cannot be
restricted by the direct searches of gauge--boson production 
discussed in Sec.~2.3

In order to reduce the number of free parameters one can make the
assumption that all blind operators affecting the Higgs
interactions have a common coupling $f$, {\it i.e.} $f = f_W =
f_B = f_{WW} = f_{BB}$ (scenario $(ii)$).  We present
in Table~\ref{tab:f} the 95\% CL allowed values of the
anomalous couplings in this scenario for different 
Higgs boson masses. In this scenario these limits 
lead to constraints on the triple gauge--boson coupling
parameters and can be compared with the constraints presented
in Eq.~(\ref{limitf:wwv}). We will go back to this comparison when
presenting the combined results from the different processes
discussed in this section.
\begin{table}[htbp]
\tcaption{95\% CL allowed range for $f/\Lambda^2$, 
from  $\gamma\gamma\gamma$ production at LEP OPAL data and 
Tevatron CDF data analysis,  from  $\gamma\gamma +  \not \!\! E_T $
Tevatron D\O ~data analysis, and from  $\gamma\gamma j j $
Tevatron D\O ~data analysis in scenario ($ii$)
We denote by --- limits worse than $|f|=200$ TeV$^{-2}$.}
\label{tab:f}
\centerline{\footnotesize\smalllineskip 
\begin{tabular}{||c||c||c||c||c||}
\hline
\hline
$m_{H}$(GeV) & \multicolumn{4}{c||}{$f/\Lambda^2$(TeV$^{-2}$)} \\
\hline 
\hline
  & $e^+ e^- \to \gamma \gamma \gamma$ & $p \bar{p} \to \gamma \gamma 
  \gamma$  & $p \bar{p} \to \gamma \gamma +  \not \!\! E_T $ &
  $p \bar{p} \to \gamma \gamma j j$\\
\hline 
\hline
100 & ( $-$64 , 57 ) & ( $-$62 , 65 ) & ( $-$28 , 57 ) & ( $-$16 , 42 )\\
\hline
120 & ( $-$82 , 70 ) & ( $-$76 , 77 ) & ( $-$37 , 62 ) & ( $-$19 , 46 )\\
\hline
140 & ( $-$192 , 175 ) & ( $-$92 , 93 ) & ( $-$48 , 72 ) & ( $-$26 , 49 )\\
\hline
160 & ( --- , --- ) & ( $-$113 , 115 ) & ( $-$62 , 84 ) & ( $-$33 , 56 )\\
\hline
180 & ( --- , --- ) & ( --- , --- ) & ( $-$103 , 123 ) & ( $-$63 , 81 )\\
\hline
200 & ( --- , --- ) & ( --- , --- ) & ( $-$160 , 164 ) & ( $-$96 , 99 )\\
\hline
220 & ( --- , --- ) & ( --- , --- ) & ( --- , --- ) & ( $-$126 , 120 ) \\
\hline
\hline
\end{tabular}}
\end{table}
As expected the limits become weaker as the Higgs becomes heavier due
to the decrease of the Higgs production cross section. 

\subsection{$p\bar p\rightarrow \gamma\gamma \,\not \!\! E_T$}
\begin{figure}[htbp]
\begin{center}
\mbox{\epsfig{file=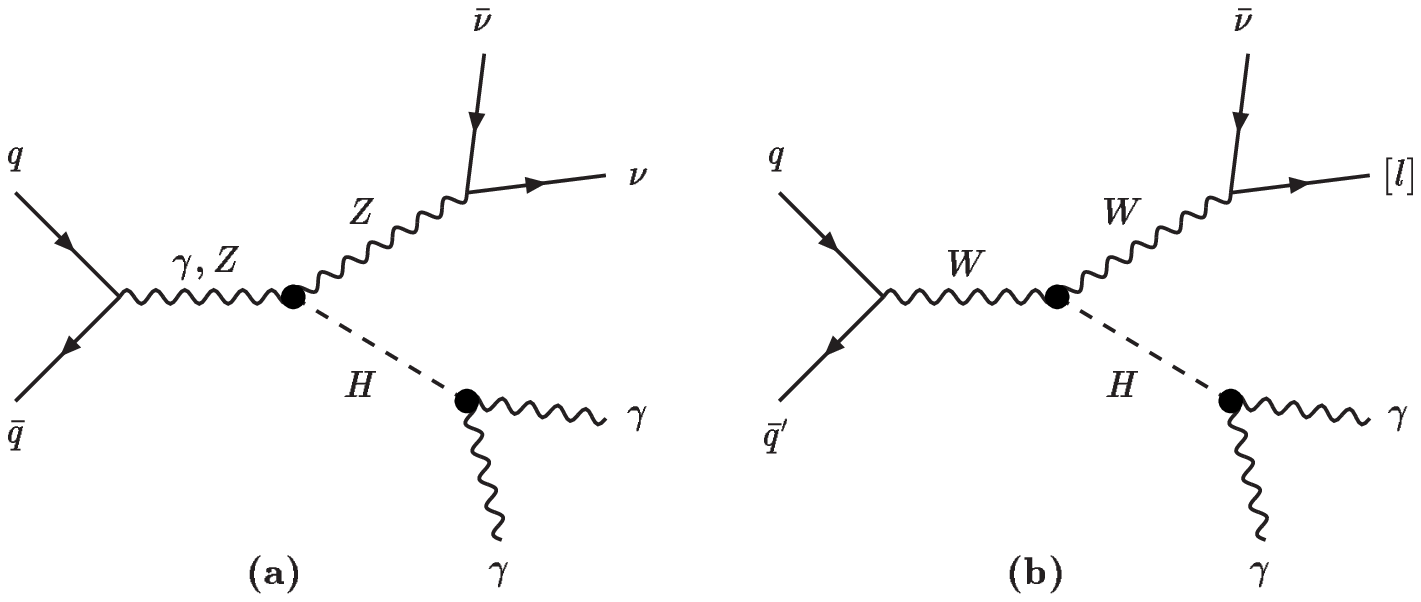,width=\textwidth}}
\end{center} 
\fcaption{Dominant anomalous contribution for 
$p\bar p\rightarrow\gamma\gamma \not \!\! E_T $ production.}
\label{fig:miss:diag} 
\end{figure}

\noindent
We examine next the process 
$p\bar p\rightarrow \gamma\gamma \,\not \!\! E_T$.
The dominant anomalous contribution to this final state when the
two photons have large invariant mass arises from 
Higgs--$W (Z)$ associated production~\cite{our:tevatronmis} as
displayed in Fig.~\ref{fig:miss:diag} where the lepton $[\ell]=e,\mu$ from
$W$ decay escapes undetected.

The D\O ~collaboration have searched for diphoton
events with large missing transverse energy in $p\bar{p}$
collisions at $\sqrt{s} = 1.8$ TeV~\cite{d0miss}.
Their analysis indicates a good  agreement with the expectations
from the Standard Model (SM). In this way, the D\O  ~Collaboration
were able to set limits on the new physics contribution to the cross section
$\sigma(p\bar{p} \to \gamma\gamma \not \!\! E_T  + X)$.

In order to compare the theoretical predictions in the presence of
the higher--dimension operators with the data collected 
by the D\O ~experiment, we must apply the same cuts of 
Ref.~\cite{d0miss}. They require that one photon has transverse energy
$E_T^{\gamma_1} > 20$ GeV and the other $E_T^{\gamma_2} > 12$
GeV, each of them with pseudorapidity in the range $|\eta^\gamma| <
1.2$ or  $1.5 < |\eta^\gamma| < 2.0$. They further require that
$\not \!\!E_T > 25$ GeV.  For the $\ell \nu \gamma \gamma$
final state, one must impose that the charged lepton is outside the
covered region of the electromagnetic calorimeter and it escapes
undetected  ($|\eta_{e}|> 2$ or $1.1<|\eta_e|<1.5 $, $
|\eta_{\mu}|> 1$). After these cuts we find that 80\% to 90\% of
the signal comes from associated Higgs--$Z$ production while 10\%
to 20\% arrises from Higgs--$W$. One must also include in the calculation
the particle identification and trigger efficiencies which vary
from 40\% to 70\% per photon~\cite{D02}. We estimate the
total effect of these efficiencies to be 35\%~\cite{our:tevatronmis}. 

Next one must consider the possible backgrounds which contribute to
the same final state. 
The main sources of background to this reaction~\cite{d0miss}
arise from SM processes containing multijets, direct photon, $W +
\gamma$, $W + j$, $Z \to ee$ and $Z \to \tau\tau \to ee$ where
photons are misidentified and/or the missing energy is
mismeasured. The D\O ~collaboration estimate the contribution of
all these backgrounds to yield $2.3 \pm 0.9$ events.  D\O
have observed two events that have passed the above
cuts in their data sample of $106.3 \pm 5.6$ pb$^{-1}$. The
invariant mass of the photon pair in these events are $50.4$, and
$264.3$ GeV. 

The results of this analysis can be used to place limits on the 
coefficients of the higher--dimension operators. Since D\O ~report that 
no event with two--photon invariant mass 
in the range $60 <M_{\gamma\gamma} < 260$ was observed, 
a $95\%$ CL in the determination of the anomalous coefficient 
$f_i$, $i=WW, BB, W, B$  is obtained by 
requiring 3 events coming only from the anomalous contributions.

\begin{figure}[htbp]
\begin{center}
\mbox{\epsfig{file=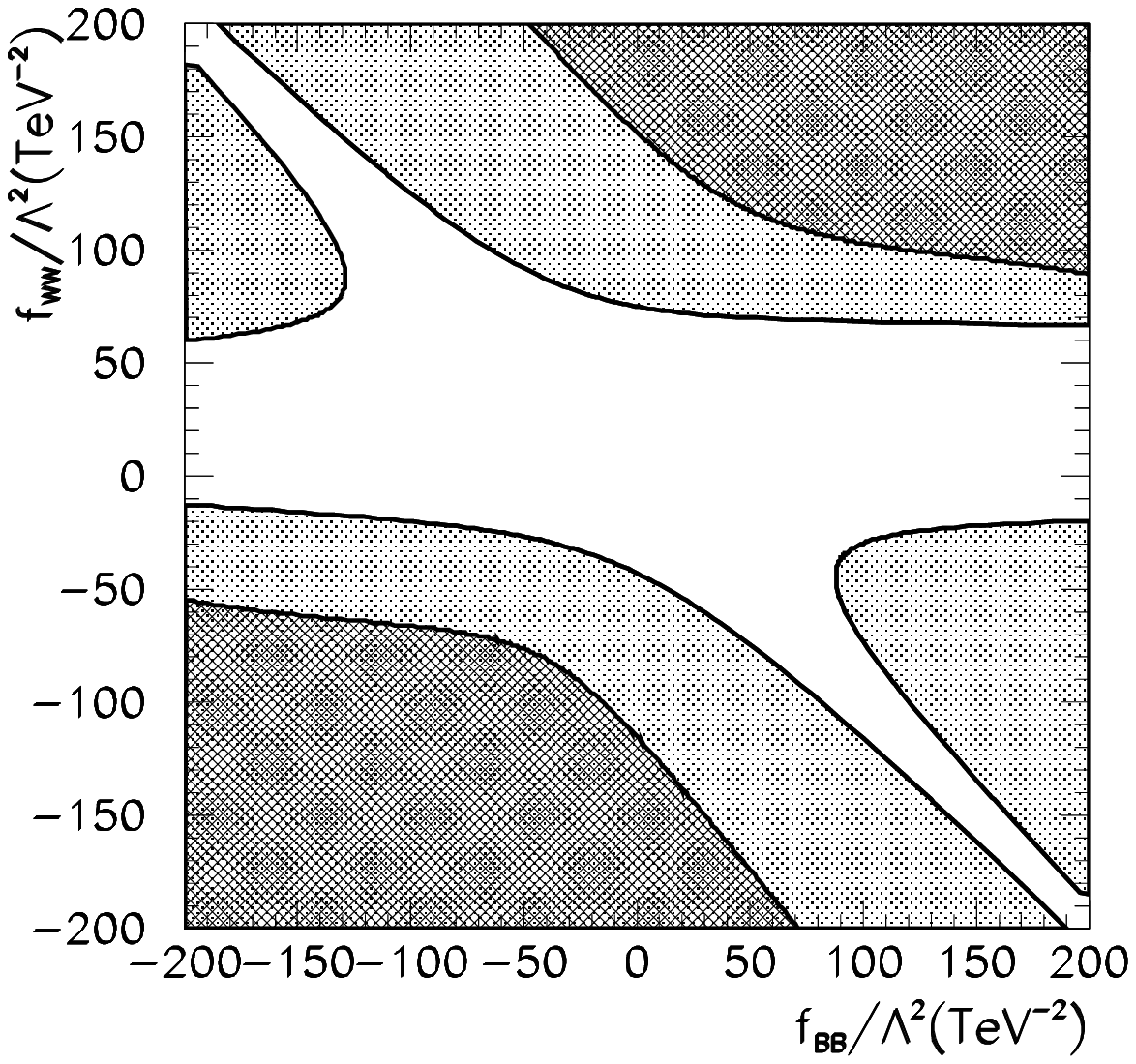,width=0.6\textwidth}}
\end{center} 
\fcaption{Excluded region at 95\% of CL in the $f_{WW} \times
f_{BB}$ plane, for an integrated luminosity of 100 pb$^{-1}$, and
for $m_{H} = 80 (140)$ GeV  [light shadow (dark shadow)].}
\label{fig:fww:fbb:miss} 
\end{figure}

In Fig.~\ref{fig:fww:fbb:miss}, we present the exclusion region in the
$f_{WW}\times f_{BB}$ plane, when we assume that just these two
coefficients are different from zero (scenario $(i))$. 
The clear (dark) shadow represents the excluded region, at $95\%$ CL, 
for $m_{H} = 80 \, (140)$ GeV. 
In Table \ref{tab:f} we give the 95\% CL allowed values of the
anomalous couplings in scenario ($ii$) {\it i.e.} $f = f_W =
f_B = f_{WW} = f_{BB}$ for different Higgs boson masses. The limits
derived from this process are weaker than those from 
$p\bar p\rightarrow jj\gamma\gamma$ due to the larger decay rate 
of $Z$ and $W$ into jets.  

\subsection{$p\bar p\rightarrow \gamma\gamma \gamma$}
\begin{figure}[htbp]
\begin{center}
\mbox{\epsfig{file=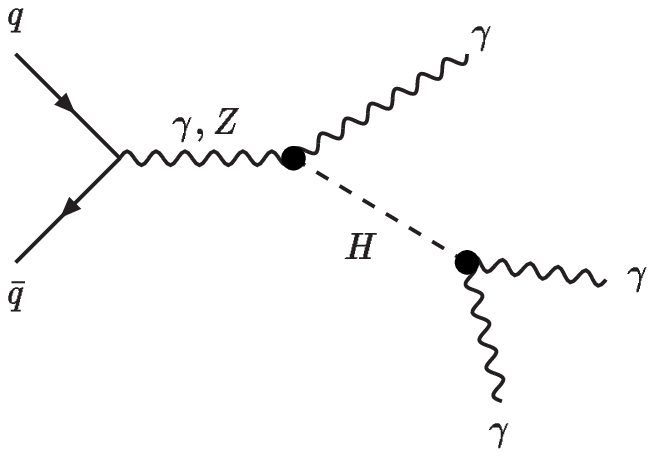,width=0.7\textwidth}}
\end{center} 
\fcaption{Dominant anomalous contribution for  
$p\bar p\rightarrow\gamma\gamma\gamma$ production.}
\label{fig:ggg:diag} 
\end{figure}

\noindent
We concentrate now in  the process 
$p\bar p\rightarrow \gamma\gamma \gamma$.
The dominant anomalous contribution to this final state for 
hard photons come from production of a Higgs boson in association
with a photon and the subsequent decay of the Higgs into 
photons~\cite{our:tevatron3a} displayed in Fig.~\ref{fig:ggg:diag}.

The CDF Collaboration~\cite{cdf} have searched for 
$\gamma\gamma\gamma$ events with two photons in
the central region of the detector ($|\eta| < 1$), with a minimum
transverse energy of 12 GeV, plus an additional photon with $E_T
> 25$ GeV. The photons were required to be separated by an angle
larger than $15^\circ$. After applying these cuts, no event was
observed, while the expected number from the background is $0.1 \pm
0.1$ in the 85 pb$^{-1}$ collected. Therefore, at 95 \% CL this
experimental result  implies that the signal should have less
than 3 events. The efficiency of identification of an isolated
photon is $68 \pm 3 \%$, for $E_T > 12$ GeV, and grows to $84 \pm
4 \%$, for $E_T > 22$ GeV. When computing the contribution from 
the higher--dimension operators to this process we must include 
the same cuts and efficiencies. 

It is important to notice that the dimension--six operators
listed in Sec.~2 do not induce $4$--point anomalous couplings
like $Z Z \gamma \gamma$, $Z \gamma\gamma \gamma$, and $\gamma
\gamma \gamma \gamma$, being these terms generated only by
dimension--eight and higher operators. Since the process 
$p\bar p\rightarrow \gamma\gamma \gamma$
involves the product of two dimension--six
operators, we should, in principle, include also in our calculations
dimension--eight operators that contribute to the above
processes at the same order in the effective Lagrangian expansion. 
Notwithstanding, we can neglect the higher--order
interactions and bound the dimension--six couplings under the
naturalness assumption that no cancelation takes place amongst
the dimension--six and --eight contributions that appear at the
same order in the expansion. 

We can now proceed and  examine which are the bounds that can
be placed on the anomalous coefficients from the negative search
of 3 photon events made by the CDF experiment. We start by
assuming that the only non--zero coefficients are the ones that
generate the anomalous $H\gamma\gamma$, {\it i.e.}, $f_{BB}$ and
$f_{WW}$ (scenario $(i)$).   
The results for the 95\% CL exclusion region in the
plane $f_{BB} \times f_{WW}$, obtained from the CDF data, are
presented in Fig.~\ref{fig:fww:fbb}. 

Finally in Table \ref{tab:f} we give the  95\% CL allowed values 
of the anomalous couplings in scenario ($ii$) {\it i.e.} $f = f_W =
f_B = f_{WW} = f_{BB}$ for different Higgs boson masses. As expected
the bounds derived from this process are weaker than the ones 
discussed in Sec.~4.2 and 4.3. This is due to the fact that 
the anomalous Higgs contribution to the $\gamma\gamma\gamma$ 
final state involves the product of two dimension--six operators
and it is therefore suppressed by $1/\Lambda^4$. 

\subsection{$e^+ e^- \rightarrow\gamma\gamma\gamma$ 
and $e^+ e^- \rightarrow\gamma\gamma+ \mbox{hadrons}$}
\begin{figure}[htbp]
\begin{center}
\mbox{\epsfig{file=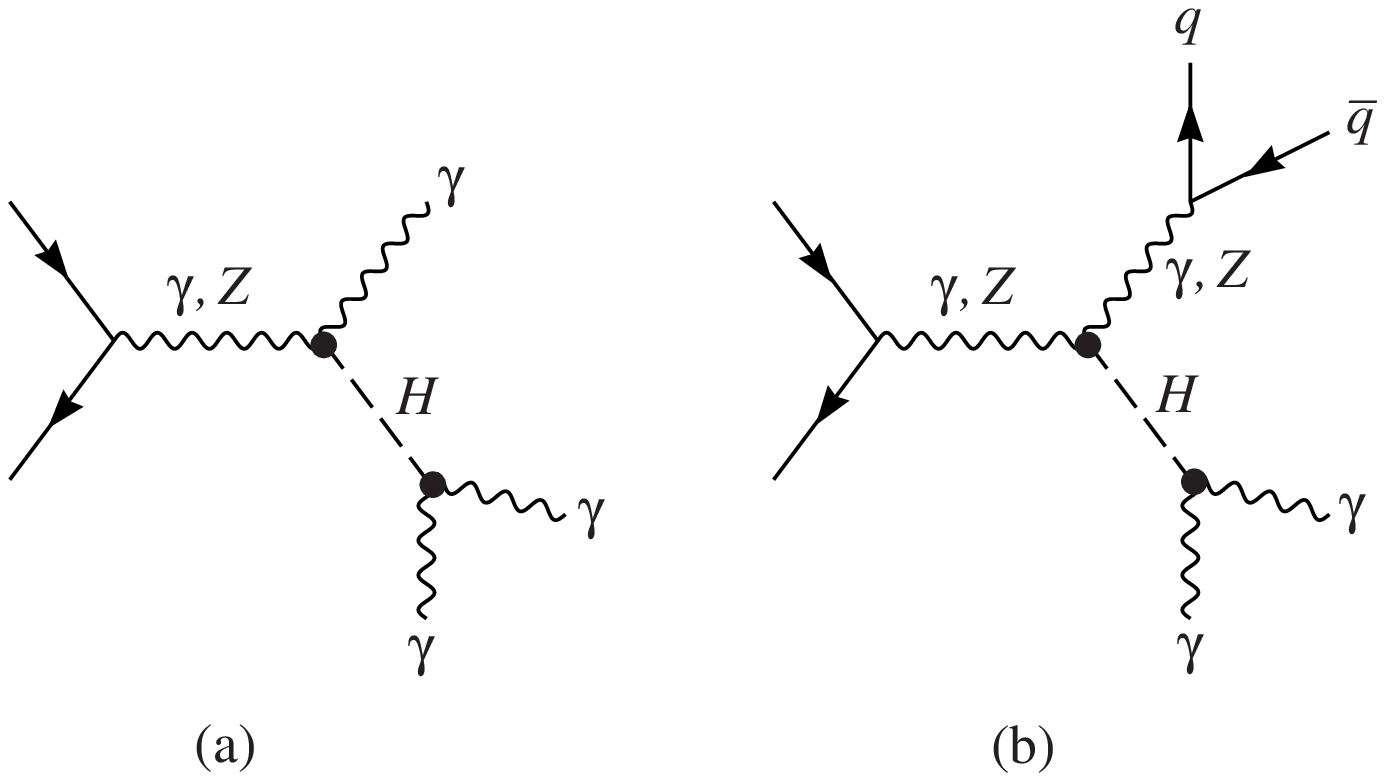,width=0.85\textwidth}}
\end{center}
\fcaption{Anomalous contribution for (a) $\gamma\gamma\gamma$
production and (b) $\gamma\gamma$ in association with hadrons.}
\label{fig:eeggg:diag}
\end{figure}

\noindent
The effect of the dimension--six operators in Higgs signatures
can also be studied at $e^+ e^-$ collisions at LEPII. Due to the
lower center--of--mass energy, the dominant contribution for 
Higgs production is expected from associated production of Higgs
 with low mass particles. We concentrate here on the processes:
\begin{eqnarray}
e^+ e^- &\rightarrow& \gamma \gamma \gamma \; ,
\label{ggg}
\\
e^+ e^- &\rightarrow& \gamma \gamma + \mbox{ hadrons} \; .
\label{ggh}
\end{eqnarray}
The Feynman diagrams describing the anomalous contributions to
the above reactions are displayed in Fig.~\ref{fig:eeggg:diag}.  
The OPAL collaboration searched for these final states. 
In Refs.~\cite{opal:gg,opal:ggg}, data taken at
several energy points in the range $\sqrt{s}=130 \,(91)$--$172$,
for the $\gamma\gamma\gamma$ ($\gamma \gamma + \mbox{ hadrons}$)
are combined in order to impose a limit on the cross section
for this process as a function of the $\gamma\gamma$ invariant mass.
In order to use OPAL results we must we also combine the expected
number of events in the presence of the anomalous operators for the 
corresponding energies and accumulated luminosities.

As for the process $p\bar p\rightarrow \gamma\gamma \gamma$, 
in $e^+ e^- \rightarrow\gamma\gamma\gamma$ 
the production and decay of the Higgs boson also involve two dimension--six
operators and we should, in principle, include in our calculations
dimension--eight operators that contribute to the above
processes. As before we neglect the higher order
interactions and bound the dimension--six couplings under the
naturalness assumption that no cancelation takes place amongst
the dimension--six and --eight contributions that appear at the
same order in the expansion. 

\begin{figure}[htbp]
\begin{center}
\mbox{\epsfig{file=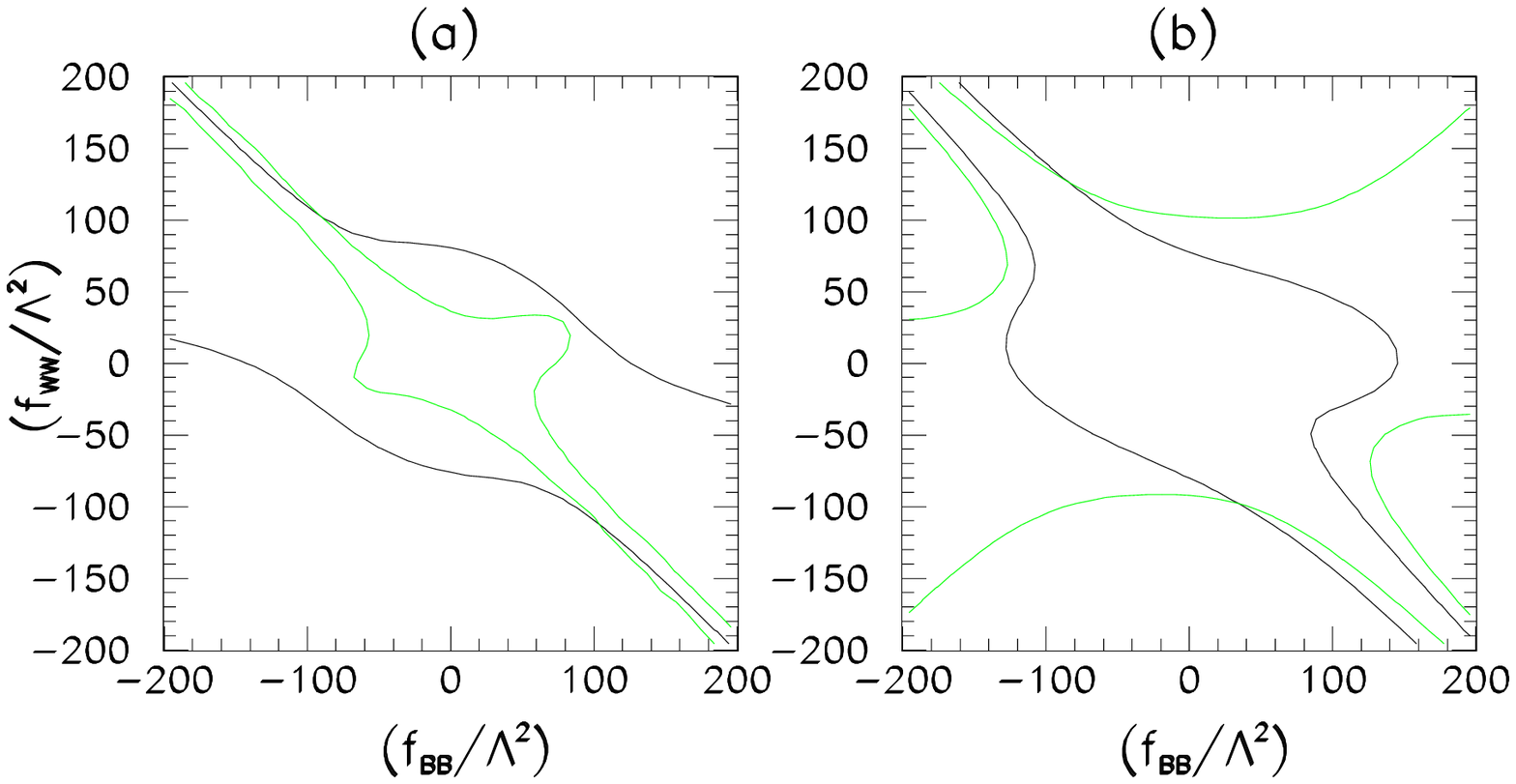,width=0.9\textwidth}}
\end{center}
\fcaption{Contour plot of $f_{BB} \times f_{WW}$, in TeV$^{-2}$.
The curves show the 95\% CL deviations from the SM total cross
section, for $e^+ e^- \to \gamma \gamma \gamma$ (dark lines) and
$e^+ e^- \to q \bar{q} \gamma \gamma$ (light lines) for (a)
$m_{H} = 60$ GeV and (b) $m_{H} = 80$ GeV. 
The excluded regions are outside the lines}
\label{fig:ggg}
\end{figure}

Assuming that the only non--zero
coefficients are the ones that generate the anomalous
$H\gamma\gamma$, {\it i.e.}, $f_{BB}$ and $f_{WW}$ (scenario $(i)$).  
we obtain the excluded region showed  in Fig.~\ref{fig:ggg}.  
For small Higgs masses
(see Fig.~\ref{fig:ggg}) the $Z$, which decays hadronically, can
be produced on mass shell and, therefore, the strongest bounds
come from the diphoton production in association with hadrons.
For higher
Higgs--boson masses ($m_{H} > 80$ GeV), the $Z$ cannot be
on--mass shell, and the $\gamma\gamma$ production accompanied by
hadrons is suppressed. In this case, only the
$\gamma\gamma\gamma$ final state is able to lead to new bounds.
Moreover, the anomalous production of a $H\gamma$ pair is also
suppressed by the phase space as $m_{H}$ increases and the limits
worsen. 
It is interesting
to notice that the bounds obtained using the above processes are
of the same order of the ones that can be extracted from the
Tevatron collider for small Higgs boson masses ($m_{H} < 80$
GeV). 

We present in Table \ref{tab:flep} the 95\% CL allowed regions of
the anomalous couplings in scenario $(ii)$. 
\begin{table}[htbp]
\tcaption{Allowed range of $f/\Lambda^2$ in TeV$^{-2}$ at 95\%
CL coming from the  processes $e^+ e^- \to \gamma \gamma
\gamma$ and $e^+ e^- \to q \bar{q} \gamma \gamma$ at LEPII.
We  assumed the scenario defined by Eq.~(\protect\ref{blind}).}
\label{tab:flep}
\centerline{\footnotesize\smalllineskip 
\begin{tabular}{||c||c||c||}
\hline
\hline
$m_{H}$(GeV) & $e^+ e^- \to \gamma \gamma \gamma$ & 
$e^+ e^- \to q \bar{q} \gamma \gamma$ \\
\hline 
\hline
60 & (  $-$56  ,   50  ) & ( $-$24  , 35  ) \\
\hline
80 & (  $-$53  ,   49  ) & ( $-$107  , 128  ) \\ 
\hline
100 & ( $-$64  ,   57  ) & ( $-$730  , 750  )\\ 
\hline
120 & ( $-$82  ,   70  ) & ------ \\ 
\hline
140 & ( $-$192  , 175  ) & ------  \\
\hline
\hline
\end{tabular}}
\end{table}
In this framework, the bounds become weaker with
the increase of the Higgs boson mass. The production of diphotons
in association with hadrons is again important only when its is
possible to produce a pair $HZ$ on mass shell. 

\subsection{Combined results: discussion}
\begin{figure}[htbp]
\begin{center}
\mbox{\psfig{file=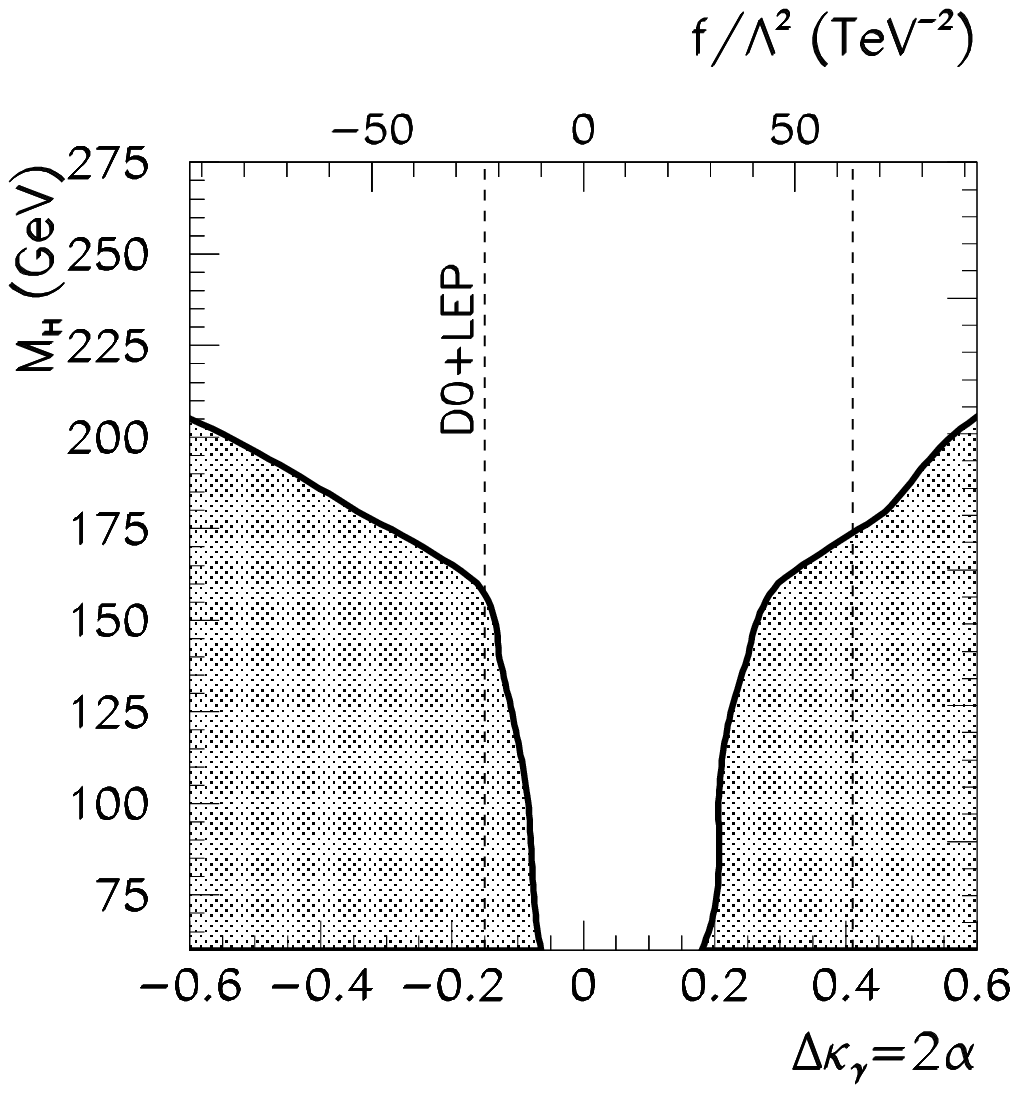,width=0.5\textwidth}}
\end{center} 
\fcaption{Excluded region in the $f \times m_{H}$
plane from the combined analysis 
from  the combined results of the  $\gamma\gamma\gamma$ 
production at LEPII, $\gamma\gamma\gamma$, $\gamma\gamma + \not \!\! E_T $, and
$\gamma\gamma j j $ production at Tevatron, assuming that all $f_i$ are equal
(see text for details).}
\label{fig:kappa}
\end{figure}

\noindent 
So far we have presented the limits on anomalous
dimension--six Higgs boson interactions that can be derived from
the study of several signatures at LEPII and Tevatron
colliders. These results obtained from the analysis of the four reactions
(\ref{proc}) can be statistically 
combined in order to obtain a better bound on the coefficient of
the effective operators (\ref{blind})~\cite{our:combined}.  
We exhibit in Fig.~\ref{fig:fww:fbb}.b the
95\% CL exclusion region in the plane $f_{BB} \times f_{WW}$
obtained from combined results in scenario $(i)$ for $m_{H}=100$ GeV.
In Fig.~\ref{fig:kappa}, we present the combined limits for the coupling 
constant $f = f_{BB}= f_{WW}= f_{B} = f_{W}=$ (scenario $(ii)$) 
for Higgs boson masses in the range of $100 \leq m_{H} \leq 220$ GeV.
In this scenario $\alpha_{W\phi}=\alpha_{B\phi}=
\alpha= \frac{m_W^2}{2\Lambda^2} f$.

We can compare now these results  with the existing limits on the 
coefficients of dimension--six operators (see Sec.~2.3)
As discussed in Sec.~2.1 for linearly realized effective 
Lagrangians, the 
modifications introduced in the Higgs  and in the vector boson sector are 
related to each other. In consequence the bounds on the new Higgs
couplings should also restrict the anomalous gauge--boson 
self--interactions.  Under the assumption of equal coefficients for all
anomalous Higgs operators, we can relate the common Higgs boson
anomalous coupling $f$ with the conventional parametrization of
the vertex $WWV$ ($V = Z^0$, $\gamma$),
\begin{eqnarray}
\Delta \kappa_\gamma    &=& \frac{m_W^2}{\Lambda^2}~ f \; ,
\nonumber \\
\Delta \kappa_Z         &=& \frac{m^2_Z}{2 \Lambda^2}~ (1 - 2 s_W^2)~ f \; ,
\label{trad} \\
\Delta g^Z_1            &=& \frac{m^2_Z}{2 \Lambda^2}~ f \; .
\nonumber
\end{eqnarray}

In Table \ref{tab:kappa}, we present the 95\% CL limit of the
anomalous coupling $\Delta\kappa_\gamma$ using the limits on $f/\Lambda^2$
obtained  through the analysis of the processes considered 
above. We also present the expected bounds 
that will be reachable at the upgraded Tevatron and at the NLC which
will be discussed next. These results
show that the present combined limit from the Higgs production analysis
is comparable with the existing bound  
from gauge--boson production (\ref{limitf:wwv}) for $m_{H} \leq 170$ GeV.
\begin{table}[htbp]
\tcaption{95\% CL allowed range for the anomalous triple gauge--boson 
couplings derived from the limits obtained for the anomalous Higgs boson 
coupling $f$.}
\label{tab:kappa}
\centerline{\footnotesize\smalllineskip 
\begin{tabular}{||c||c||c||}
\hline
\hline
Process & $m_{H}$ (GeV) & $ \Delta\kappa_\gamma=2\,\alpha= 2\,\alpha_{B \Phi} = 
2,\alpha_{W \Phi}$ \\ 
\hline 
\hline
Combined Tevatron RunI + LEPII & 100 &( $-$0.084 , 0.204 ) \\
\hline
Combined Tevatron RunII & 100 & ( $-$0.048 , 0.0122 ) \\
\hline
Combined Tevatron TeV33 & 100 & ( $-$0.020 , 0.036 ) \\
\hline
$e^+ e^- \to W^+ W^- \gamma$ at NLC & 200 & ( $-$0.020 , 0.026 ) \\
\hline
$e^+ e^- \to Z^0 Z^0 \gamma$ at NLC & 200 & ( $-$0.016 , 0.024 ) \\
\hline
\hline
\end{tabular}}
\end{table}

\subsection{Future perspectives}
\noindent 
The effect of the anomalous operators becomes more evident with
the increase of energy, and higher sensitive to smaller values of the
anomalous coefficients can be achieved by studying their contribution to
different processes at the upgraded Tevatron collider or at new machines,
like the Next Linear Collider. 

The analysis of the reactions $p \bar{p} \to\gamma \gamma \not \!\! E_T$ 
and $p \bar{p} \to \gamma \gamma j j$ presented before can be repeated for 
the upgraded  Tevatron collider. We consider here that the upgraded 
Tevatron Run II will collect an integrated luminosity of 1
fb$^{-1}$, and  TeV33 will deliver  10 fb$^{-1}$~\cite{tevatron}.
In the results presented next we will assume the same cuts and detection 
efficiencies as given in the previous sections. 

For the $\gamma\gamma\gamma$ final state it is possible to 
improve the sensitivity to the anomalous coefficients
by implementing additional kinematical cuts~\cite{our:tevatron3a}.
Best results are obtained for the following set of cuts:
$E_{T_{1}} > 40$ GeV, with $E_{T_{2,3}} > 12$ GeV where 
the three photons have been ordered according to their transverse energy,
{\it i.e.\/} $E_{T_{1}} > E_{T_{2}} > E_{T_{3}}$.
The photons are always required to be in the central region of the detector
($|\eta_{i}| < 1$) where there is sensitivity for electromagnetic
showering. In the estimates presented here  the same detection
efficiency for photons as considered by the CDF Collaboration is 
assumed~\cite{cdf}.

Table \ref{tab:run2} contains the 95\% CL 
limits on the anomalous couplings that could be achievable at 
Tevatron Run II and at TeV33 
for each individual process. All couplings are assumed equal 
(scenario $(ii)$)
and the Higgs boson mass is varied in the range $100 \leq 
m_{H} \leq 220$ GeV.   
The combination of the results obtained from the analysis of the three 
reactions leads to the
improved bounds given in  
Table \ref{tab:run2:comb}. Comparing these results with
those in Fig.~\ref{fig:kappa} we observe an expected improvement of about 
a factor $\sim 2$--$3$ [$\sim 4$-$-6$] for the combined limits 
at RunII [TeV33].  
\begin{table} [htbp]
\tcaption{95\% CL allowed range for $f/\Lambda^2$, 
from  $\gamma\gamma\gamma$, $\gamma\gamma + \not \!\! E_T $, 
$\gamma\gamma j j $ production at Tevatron Run II [TeV33] assuming  
all $f_i$ to be equal. 
We denote by --- limits worse than $|f|=100$ TeV$^-2$.}
\label{tab:run2}
\centerline{\footnotesize\smalllineskip 
\begin{tabular}{||c||c||c||c||}
\hline
\hline
$m_{H}$(GeV) & \multicolumn{3}{c||}{$f/\Lambda^2$(TeV$^{-2}$)} \\
\hline 
\hline
  & $p \bar{p} \to \gamma \gamma \gamma$  & $p \bar{p} 
  \to \gamma \gamma +  \not \!\! E_T $  &
  $p \bar{p} \to \gamma \gamma j j$  \\
\hline 
\hline
100 & ( $-$24 , 24 ) [ $-$13 , 15 ] 
& ( $-$16 , 36 ) [ $-$9.4 , 26 ] 
& ( $-$9.2 , 22 ) [ $-$3.3 , 5.6 ] 
\\
\hline
120 & ( $-$26 , 26 ) [$-$14 , 14 ] 
& ( $-$20 , 39 )  [ $-$15 , 27 ]
& ( $-$8.6 , 21 ) [ $-$3.4 , 5.9 ] 
\\
\hline
140 & ( $-$30 , 31 )[ $-$15 , 16] 
& ( $-$25 , 44 ) [ $-$14 , 30]
& ( $-$10 , 23 )  [ $-$4.5 , 8.9]
\\
\hline
160 & ( $-$36 , 38 ) [$-$17 , 19] 
& ( $-$29 , 50 )  [$-$14 , 33]
& ( $-$11 , 24 )  [$-$6.0 , 14]
\\
\hline
180 & ( --- , --- )[ --- , --- ] 
& ( $-$63 , 72 )   [ $-$46 , 53 ]
& ( $-$26 , 34 )   [ $-$16 , 24 ]
\\
\hline
200 & ( --- , --- ) [ --- , --- ]   
& ( $-$87 , 90 ) [$-$50 , 53]
& ( $-$33 , 40 ) [ $-$17 , 23]
\\
\hline
220 & ( --- , --- ) [ --- , --- ]
& ( --- , --- )  [--- , --- ]
& ( $-$42 , 45 ) [$-$19 , 26]
\\
\hline
\hline 
\end{tabular}}
\end{table}
\begin{table} [htbp]
\tcaption{95\% CL allowed range for $f/\Lambda^2$, 
from  the combinations of $\gamma\gamma\gamma$, $\gamma\gamma +  
\not \!\! E_T $, 
$\gamma\gamma j j $ production at Tevatron Run II [TeV33] assuming  
all $f_i$ to be equal. }
\label{tab:run2:comb}
\centerline{\footnotesize\smalllineskip 
\begin{tabular}{||c||c||}
\hline
\hline
$m_{H}$(GeV) & $f/\Lambda^2$(TeV$^{-2}$) \\
\hline 
\hline
& COMBINED\\
\hline 
\hline
100 & ( $-$7.6 , 19 )[ $-$3 , 5.6 ]
\\
\hline
120 & ( $-$7.4 , 18 )[$-$3.3 , 5.9]
\\
\hline
140 & ( $-$9.1 , 20 )[ $-$4.0 , 8.7]
\\
\hline
160 & ( $-$9.9 ,22 )  [$-$5.1 , 13]
\\
\hline
180 & ( $-$24 , 33 )   [ $-$16 , 24 ]  \\
\hline
200 & ( $-$32 , 39 ) [ $-$17 , 23 ] \\
\hline
220 & ( $-$42 , 45 ) [$-$19 , 26 ]  \\
\hline
\hline 
\end{tabular}}
\end{table}

The Next Linear electron--positron Collider will open an
important opportunity to further improve the search for new
physics. In particular, the anomalous Higgs boson couplings can
be investigated in the processes~\cite{our:NLCWWA,our:NLCZZA}:
\begin{eqnarray}
e^+ e^- &\to& W^+ W^- \gamma \label{wwg}\; , \\
e^+ e^- &\to& Z^0 Z^0 \gamma  \label{zzg}\; .
\end{eqnarray}

We discuss here the sensitivity of NLC to these processes assuming a
center--of--mass energy of  $\sqrt{s} = 500 \; \mbox{GeV}$
and an integrated luminosity  ${\cal L} = 50  \;\mbox{fb}^{-1}$.
In order to account for standard detector effects 
a cut in the photon  energy of $E_\gamma > 20 \;
\mbox{GeV}$ was adopted and the angle between  any two particles was required
to be larger than $15^\circ$. 

One can investigate the  different distributions of the final
state particles in order to search for kinematical cuts that could
improve the NLC sensitivity. The most promising variable is the
photon transverse momentum as the
contribution of the anomalous couplings is larger in the high
$p_{T_\gamma}$ region. This is understood because 
the anomalous signal is dominated by
on--mass--shell Higgs--$\gamma$ production with the subsequent $H \to W^+
W^-$ or $Z^0Z^0$ decay and
the photon transverse momentum is distributed around 
the monochromatic peak
$E_\gamma^{\mbox{mono}}= (s - m_{H}^2)/(2 \sqrt{s})$. In consequence 
for Higgs boson masses in the range $2 m_{W,Z} \leq m_{H} 
\leq (\sqrt{s} -
E_\gamma^{\mbox{min}})$ GeV, where on--shell production is allowed,  
a cut of $p_{T_\gamma} > 100$ drastically reduces the background.  
For lighter Higgs bosons, {\it e.g.\/} $m_{H} <  2
m_{W,Z}$, the $p_{T_\gamma}$ cut is ineffective since the Higgs
boson is off--mass--shell and the peak in the photon transverse
momentum distribution disappears. This makes the attainable 
bounds on the anomalous coefficients that could be obtained 
from the $W^+ W^- (Z^0Z^0) \gamma$ production to be very loose.

\begin{figure}
\begin{center}
\mbox{\epsfig{file=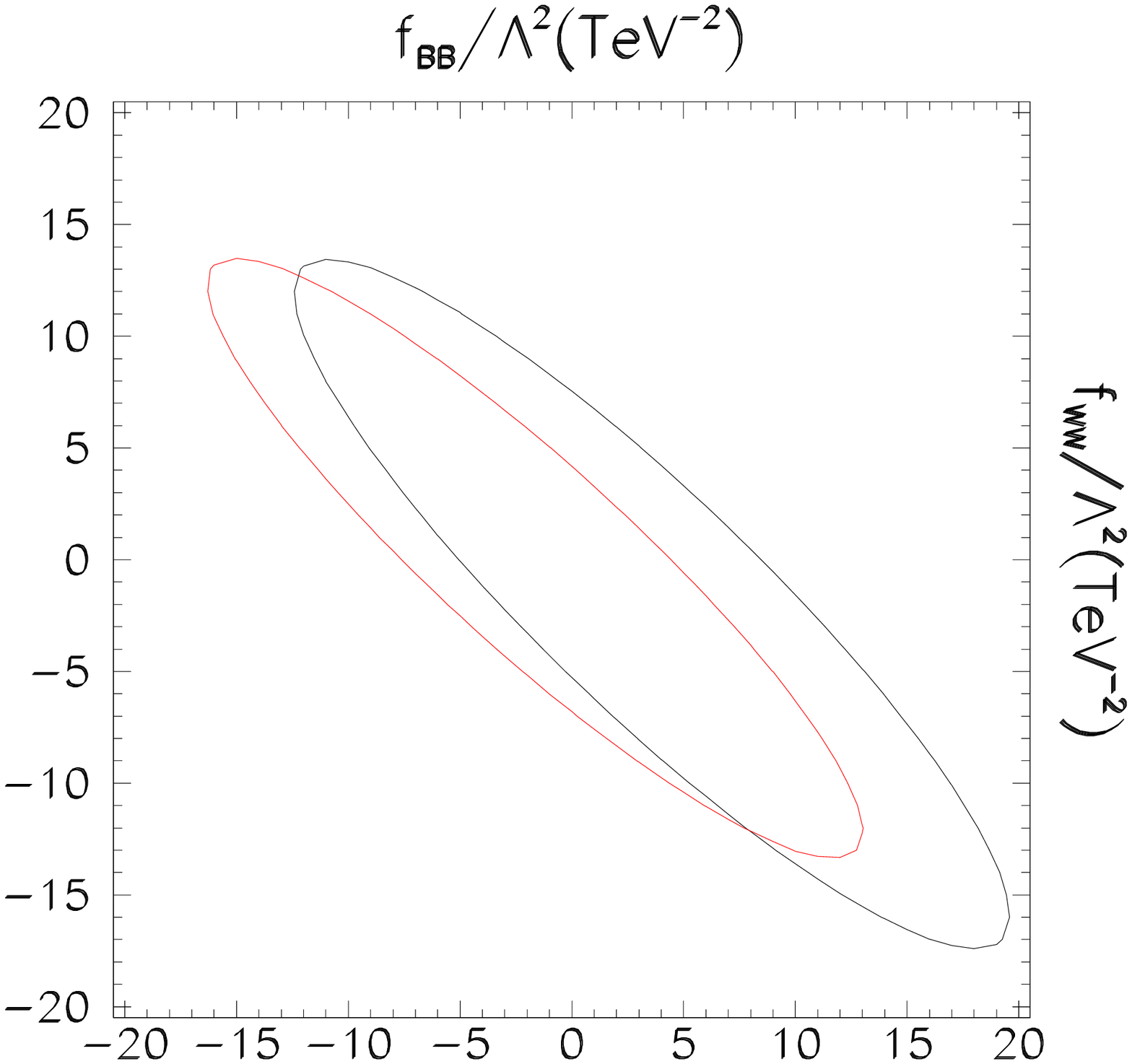,width=0.5\textwidth}}
\end{center}
\fcaption{Contour plot of $f_{BB} \times f_{WW}$, from $e^+ e^-
\to W^+ W^- \gamma$ (black line) and $e^+ e^- \to Z^0 Z^0 \gamma$ 
(red line) at NLC, for $m_{H} =200$ GeV with a cut of $p_{T_\gamma} 
> 100$ GeV. The curves show the 95\% CL deviations from the SM 
total cross section.}
\label{fig:fww:fbb:nlc}
\end{figure}

In  Fig.~\ref{fig:fww:fbb:nlc} 
we show the region in the plane $f_{BB} 
\times f_{WW}$ for $m_{H}=200$ GeV that could be excluded at
95\% CL from the study of reactions (\ref{wwg}) and 
(\ref{zzg}). Notice that for these two reactions the 
exclusion region closes the gap at $f_{BB} = -
f_{WW}$ since the anomalous decay widths $H \to W^+W^- (Z^0 Z^0)$
do not vanish along this axis as we have seen in Sec.~3.2. 

We present in Table \ref{tab:fnlc} the sensitivity to the coefficient
$f/\Lambda^2$ based on a 95\% CL deviation in the total
cross  section for a Higgs mass in the range $170  \leq m_{H} \leq
350$ GeV in scenario ($ii$). The results coming from the $Z^0 Z^0\gamma$
production are a little better than the ones obtained from $W^+
W^- \gamma$ production, and they can improve by one order of magnitude
the actual limits derived from LEP and Tevatron data analyses.
\begin{table}
\tcaption{95\% CL allowed range for $f/\Lambda^2$, 
from $W^+ W^- \gamma$ and 
$Z^0 Z^0 \gamma$ production at NLC, assuming all $f_i$ 
to be equal.}
\label{tab:fnlc}
\centerline{\footnotesize\smalllineskip \begin{tabular}{||c||c||c||}
\hline
\hline
$m_{H}$(GeV) & \multicolumn{2}{c||}{$f/\Lambda^2$(TeV$^{-2}$)} \\
\hline 
\hline
  & {$e^+ e^- \to W^+ W^- \gamma$ at NLC} & 
  {$e^+ e^- \to Z^0 Z^0 \gamma$ at NLC} \\
\hline 
\hline
170 & ( $-$2.3 , 3.7 ) & ( --- , --- ) \\
\hline
200 & ( $-$3.2 , 4.0 ) & ( $-$2.6 ,3.9  ) \\
\hline
250 & ( $-$4.3 , 4.8 ) & ( $-$3.2 , 4.3 ) \\
\hline
300 & ( $-$6.3 , 6.3 ) & ( $-$4.7 , 5.2 ) \\
\hline
350 & ( $-$12 , 9.5 ) & ( $-$7.1 , 8.3 )\\
\hline
\hline
\end{tabular}}
\end{table}

\section{Conclusions}
\noindent 
A consistent description of 
the effect of new physics in the bosonic sector of the SM in terms
of effective Lagrangians, implies the presence of anomalous Higgs 
couplings to the gauge bosons. In this review we have concentrated
on the effects of these new interactions on the Higgs boson 
phenomenology. 

In the effective Lagrangian language, we have described the effects of
the new physics at low energy by including in the Lagrangian
higher--dimension operators.  In building these operators we have used
a linear realization of the $SU(2)_L \times U(1)_Y$ gauge symmetry and
we have included all operators $C$ and $P$ invariant constructed out
of the gauge--boson and the Higgs fields while keeping the fermionic
sector unchanged.  Such effective Lagrangian contains eleven
dimension-six operators with unknown coefficients.  Four of these
operators, ${\cal O}_{\Phi,1}$, ${\cal O}_{DW}$, ${\cal O}_{DB}$, and
${\cal O}_{BW}$, modify the gauge--boson two--point functions at tree
level while three operators, ${\cal O}_{WWW}$, ${\cal O}_{W}$, and
${\cal O}_{B}$, enter at lower order in the gauge--boson three--point
functions.  We have summarized the constraints on these operators
arising from their contributions to existing low--energy observables
as well as to the direct gauge--boson production at the Tevatron
Collider and LEPII.

The operators ${\cal O}_{WW}$, ${\cal O}_{BB}$, 
${\cal O}_{W}$, and  ${\cal O}_{B}$,     
enter at lower order in the Higgs couplings to the gauge bosons.  
Most of this review is concentrated on the study of the effects of
these operators on the Higgs boson signatures, and on the possibility
of constraining their coefficients by using the negative results of 
searches at LEPII and the Tevatron colliders.
One of the most interesting features associated with the presence 
of these operators is the enhancement of the Higgs decay rate in
two photons what makes the Higgs searches particularly clean 
at hadron colliders. We have shown how the use of the results from searches 
of final states containing photons such as :
$p\, \bar p  \rightarrow  j \, j \, \gamma\, \gamma$, 
$p\, \bar p \rightarrow \gamma \,\gamma \,+ \,\not \!\! E_T $,
$p\, \bar p \rightarrow \gamma\, \gamma\, \gamma$ and 
$e^+\, e^-  \rightarrow \gamma\, \gamma\, \gamma$ 
at the Tevatron and LEPII can be used to place 
limits on the values of the coefficients of the higher--dimension
operators, or, in other words, on the scale of new physics.  
Since we have  concentrated on the effects associated with 
the Higgs decay into photons which involve the operators 
${\cal O}_{WW}$ and ${\cal O}_{BB}$,  the bounds apply 
undoubtedly to the coefficients of those operators. 
The limits on these coefficients are summarized in 
Fig.~\ref{fig:fww:fbb}. 

The operators ${\cal O}_{W}$, and  ${\cal O}_{B}$ modify
both the Higgs production rates as well as the gauge 
boson self--couplings. We have discussed how if
we further assume that the coefficients of the 
four operators ${\cal O}_{WW}$ and ${\cal O}_{BB}$, 
${\cal O}_{W}$, and  ${\cal O}_{B}$ are equal, both 
effects can be compared. 
The limits obtained from Higgs boson searches under this assumption 
are summarized 
in Tables \ref{tab:f}, \ref{tab:kappa}, and in Fig.~\ref{fig:kappa}.  
Under the assumption of equal coefficients for all
anomalous Higgs operators, we can relate the common Higgs boson
anomalous coupling $f$ with the conventional parametrization of
the vertex $WWV$ with   
$\Delta \kappa_\gamma=m_W^2/\Lambda^2 ~f$. 
The present combined limit from the Higgs production analysis
is comparable with the existing bound  
from gauge--boson production for $m_{H} \leq 170$ GeV.  
We have also discussed  how
the sensitivity to anomalous Higgs couplings  can be expected to 
improve by a factor $2$--$6$ 
at future Tevatron runs and by about one order of magnitude at the
the NLC and it will reach close to the strong bounds on the four 
``not-blind'' operators that contribute to the four--fermion amplitudes 
at tree level.

\nonumsection{Acknowledgements} 
\noindent 
I wish to thank all my collaborators in the subject of anomalous
higgs couplings: F. de Campos, O.J.P. Eboli, S. Lietti, S.F. Novaes, 
and R. Rosenfeld. I also want to thank R. Vazquez for his careful 
reading of the manuscript and for many useful comments and suggestions.  
It is also a pleasure to acknowledge
the Instituto de Fisica Teorica de la Universidade Estadual 
Paulista in Sao Paulo, Brazil,  for their warm hospitality during 
my stays there when most of the results included in this review were 
obtained. This work was supported by
grants CICYT AEN96-1718, DGICYT PB95-1077 and  DGICYT PB97-1261, 
by the EEC under the TMR contract ERBFMRX-CT96-0090 and 
by Funda\c{c}\~ao de Amparo \`a Pesquisa do Estado de S\~ao Paulo
(FAPESP).

\nonumsection{References}

\appendix 

Here we give the Feynman rules for the triple and quartic 
vertices from operators listed in Sec.~2.1
We do not include the contributions from ``not--blind'' operators
(\ref{noblind}).

The couplings of the Higgs to gauge bosons are:
\begin{center} 
\begin{picture}(300,100)(30,0)
\DashArrowLine(0,60)(40,60){4}
\Text(-5,60)[r]{$H$}
\Text(25,50)[r]{$k$}
\Photon(40,60)(80,90){2}{5}
\Text(60,85)[r]{$k_1$}
\Text(60,37)[r]{$k_2$}
\Text(100,90)[r]{$A_\alpha$}
\Text(100,30)[r]{$A_\beta$}
\Photon(40,60)(80,30){2}{5}
\ArrowLine(60,75)(55,75)
\ArrowLine(60,45)(55,50)
\Text(120,60)[l]{$i\; \frac{\displaystyle g m_W}{\displaystyle \Lambda^2} 2 s^2 (f_{BB} + f_{WW})[g^{\alpha \beta} (k_1 \cdot k_2) - k_1^\beta k_2^\alpha ]$}
\end{picture}
\end{center}
\begin{center}
\begin{picture}(300,100)(30,0)
\DashArrowLine(0,60)(40,60){4}
\Text(-5,60)[r]{$H$}
\Text(25,50)[r]{$k$}
\Photon(40,60)(80,90){2}{5}
\Text(60,85)[r]{$k_1$}
\Text(60,37)[r]{$k_2$}
\Text(100,90)[r]{$A_\alpha$}
\Text(100,30)[r]{$Z_\beta$}
\Photon(40,60)(80,30){2}{5}
\ArrowLine(60,75)(55,75)
\ArrowLine(60,45)(55,50)
\Text(120,60)[l]{$i \;\frac{\displaystyle g m_W}{\displaystyle 
\Lambda^2} \frac{\displaystyle s}{\displaystyle c}
\Big\{2(s^2 f_{BB}-c^2 f_{WW}) [-g^{\alpha \beta}(k_1 \cdot k_2)+ 
k_2^\alpha k_1^\beta] +$}
\Text(125,40)[l]{$\frac{\displaystyle 1}{\displaystyle 2}(f_W -f_B)
[-g^{\alpha \beta}(k_1^2 + k_1 \cdot k_2)+
(k_1^\alpha + k_2^\alpha) k_1^\beta]\Big\}$}
\end{picture}
\end{center}
\begin{center}
\begin{picture}(300,100)(30,0)
\DashArrowLine(0,60)(40,60){4}
\Text(-5,60)[r]{$H$}
\Text(25,50)[r]{$k$}
\Photon(40,60)(80,90){2}{5}
\Text(60,85)[r]{$k_1$}
\Text(60,37)[r]{$k_2$}
\Text(100,90)[r]{$Z_\alpha$}
\Text(100,30)[r]{$Z_\beta$}
\Photon(40,60)(80,30){2}{5}
\ArrowLine(60,75)(55,75)
\ArrowLine(60,45)(55,50)
\Text(120,70)[l]{$i \;\frac{\displaystyle g m_W}{\displaystyle 
 \Lambda^2} \frac{\displaystyle 1}{\displaystyle 2c^2}
\Big\{4(s^4f_{BB}+c^4f_{WW})
[g^{\alpha \beta}(k_1 \cdot k_2) - k_2^\alpha 
k_1^\beta]  + $}
\Text(125,50)[l]{$
(c^2f_W+s^2f_B)[-g^{\alpha \beta}(k_1^2 +k_2^2 +2 k_1 \cdot k_2) +
(k_1^\alpha k_1^\beta + 2 k_2^\alpha k_1^\beta + k_2^\alpha k_2^\beta )]
\Big\}$} 
\end{picture}
\end{center}
\vskip 1cm
\begin{center}
\begin{picture}(300,100)(30,0)
\DashArrowLine(0,60)(40,60){4}
\Text(-5,60)[r]{$H$}
\Text(25,50)[r]{$k$}
\Photon(40,60)(80,90){2}{5}
\Text(60,85)[r]{$k_1$}
\Text(60,37)[r]{$k_2$}
\Text(100,90)[r]{$W^+_\alpha$}
\Text(100,30)[r]{$W^-_\beta$}
\Photon(40,60)(80,30){2}{5}
\ArrowLine(60,75)(55,75)
\ArrowLine(60,45)(55,50)
\Text(120,70)[l]{$i \;\frac{\displaystyle g m_W}{\displaystyle 
\Lambda^2} \Big\{\frac{\displaystyle f_W}{\displaystyle 2}
[(k_1^\alpha k_1^\beta + k_2^\alpha k_2^\beta) - 
g^{\alpha \beta} (k_1^2 + k_2^2)] + $}
\Text(125,50)[l]{$
(f_W-2f_{WW})[k_2^\alpha k_1^\beta - g^{\alpha \beta}
(k_1 \cdot k_2)] \Big\}$}
\end{picture}
\end{center}
The triple vector boson self--couplings: 
\begin{center}
\begin{picture}(300,140)(30,0)
\Photon(0,60)(40,60){2}{5}
\ArrowLine(18,60)(23,60)
\Text(-5,60)[r]{$A^\alpha$}
\Text(25,50)[r]{$k_1$}
\Photon(40,60)(80,90){2}{5}
\Text(60,85)[r]{$k_2$}
\Text(60,37)[r]{$k_3$}
\Text(100,90)[r]{$W^+_\beta$}
\Text(100,30)[r]{$W^-_\gamma$}
\Photon(40,60)(80,30){2}{5}
\ArrowLine(60,75)(55,75)
\ArrowLine(60,45)(55,50)
\Text(120,80)[l]{$i \;\frac{\displaystyle g s}{\displaystyle 
2 \Lambda^2} 
\Big\{m_W^2(f_{B}+f_{W})
[g^{\alpha \beta} k_1^\gamma- g^{\alpha \gamma} k_1^\beta]+ $}
\Text(125,60)[l]{$-3 g^2 f_{WWW}\Big[k_1^\beta k_2^\gamma k_3^\alpha-
k_1^\gamma k_2^\alpha k_3^\beta+(k_1 \cdot k_2)
\left(g^{\alpha\gamma}k_3^\beta -g^{\beta\gamma}k_3^\alpha \right)+ $}
\Text(125,40)[l]{$(k_1 \cdot k_3)
\left(g^{\beta\gamma}k_2^\alpha -g^{\alpha\beta}k_2^\gamma \right)
+(k_2 \cdot k_3)
\left(g^{\alpha\beta}k_1^\gamma - g^{\alpha \gamma} k_1^\beta\right)\Big]
\Big\}$}
\end{picture}
\vskip 1cm
\end{center}
\begin{center}
\begin{picture}(300,160)(30,-30)
\Photon(0,60)(40,60){2}{5}
\ArrowLine(20,60)(25,60)\Text(-5,60)[r]{$Z^\alpha$}
\Text(25,50)[r]{$k_1$}
\Photon(40,60)(80,90){2}{5}
\Text(60,85)[r]{$k_2$}
\Text(60,37)[r]{$k_3$}
\Text(100,90)[r]{$W^+_\beta$}
\Text(100,30)[r]{$W^-_\gamma$}
\Photon(40,60)(80,30){2}{5}
\ArrowLine(60,75)(55,75)
\ArrowLine(60,45)(55,50)
\Text(120,90)[l]{$i \;\frac{\displaystyle g }{\displaystyle 
2 \Lambda^2 c} 
\Big\{- m_W^2 s^2(f_{B}+f_{W})
[g^{\alpha \beta} k_1^\gamma- g^{\alpha \gamma} k_1^\beta]+ $}
\Text(125,70)[l]{$ m_W^2 f_W
[g^{\alpha \gamma} (k_3-k_1)^\beta+ 
g^{\alpha \beta} (k_1-k_2)^\gamma+ g^{\beta\gamma} (k_2-k_3)^\alpha]+ $}
\Text(125,50)[l]{$-3 g^2 c^2 f_{WWW}\Big[k_1^\beta k_2^\gamma k_3^\alpha-
k_1^\gamma k_2^\alpha k_3^\beta+(k_1 \cdot k_2)
\left(g^{\alpha\gamma}k_3^\beta -g^{\beta\gamma}k_3^\alpha \right)+ $}
\Text(125,30)[l]{$(k_1 \cdot k_3)
\left(g^{\beta\gamma}k_2^\alpha -g^{\alpha\beta}k_2^\gamma \right)
+(k_2 \cdot k_3)
\left(g^{\alpha\beta}k_1^\gamma - g^{\alpha \gamma} k_1^\beta\right)\Big]
\Big\}$}
\end{picture}
\end{center}
\vskip 1cm

The quartic couplings for the gauge bosons, involve the couplings 
$f_W$, $f_{WWW}$ and $f_{DW}$. We give here the expressions for
$f_{WWW}=f_{DW}=0$ which are the relevant ones for our study:
\begin{center}
\begin{picture}(300,100)(30,0)
\Photon(40,60)(0,30){2}{5}
\Photon(40,60)(0,90){2}{5}
\Text(28,85)[r]{$k_1$}
\Text(28,37)[r]{$k_2$}
\Text(-5,90)[r]{$Z^\alpha$}
\Text(-5,30)[r]{$Z^\beta$}
\ArrowLine(20,75)(25,70)
\ArrowLine(20,45)(25,50)
\Photon(40,60)(80,90){2}{5}
\Text(60,85)[r]{$k_3$}
\Text(60,37)[r]{$k_4$}
\Text(100,90)[r]{$W^+_\gamma$}
\Text(100,30)[r]{$W^-_\rho$}
\Photon(40,60)(80,30){2}{5}
\ArrowLine(60,75)(55,75)
\ArrowLine(60,45)(55,50)
\Text(120,60)[l]{$i \;\frac{\displaystyle g^2 m_W^2 }{\displaystyle 
\Lambda^2} f_W\Big[ g^{\alpha\gamma} g^{\beta\rho}+ 
g^{\alpha\rho} g^{\beta\gamma}- 2g^{\alpha\beta} g^{\gamma\rho}\Big]$}
\end{picture}
\end{center}
\begin{center}
\begin{picture}(300,100)(30,0)
\Photon(40,60)(0,30){2}{5}
\Photon(40,60)(0,90){2}{5}
\Text(28,85)[r]{$k_1$}
\Text(28,37)[r]{$k_2$}
\Text(-5,90)[r]{$A^\alpha$}
\Text(-5,30)[r]{$Z^\beta$}
\ArrowLine(20,75)(25,70)
\ArrowLine(20,45)(25,50)
\Photon(40,60)(80,90){2}{5}
\Text(60,85)[r]{$k_3$}
\Text(60,37)[r]{$k_4$}
\Text(100,90)[r]{$W^+_\gamma$}
\Text(100,30)[r]{$W^-_\rho$}
\Photon(40,60)(80,30){2}{5}
\ArrowLine(60,75)(55,75)
\ArrowLine(60,45)(55,50)
\Text(120,60)[l]{$i \;\frac{\displaystyle g^2 m_W^2 s }{\displaystyle 
\Lambda^2 c} f_W\Big[ g^{\alpha\gamma} g^{\beta\rho}+ 
g^{\alpha\rho} g^{\beta\gamma}- 2g^{\alpha\beta} g^{\gamma\rho}\Big]$}
\end{picture}
\end{center}
\begin{center}
\begin{picture}(300,100)(30,0)
\Photon(40,60)(0,30){2}{5}
\Photon(40,60)(0,90){2}{5}
\Text(28,85)[r]{$k_1$}
\Text(28,37)[r]{$k_2$}
\Text(-5,90)[r]{$W^+_\alpha$}
\Text(-5,30)[r]{$W^-_\beta$}
\ArrowLine(20,75)(25,70)
\ArrowLine(20,45)(25,50)
\Photon(40,60)(80,90){2}{5}
\Text(60,85)[r]{$k_3$}
\Text(60,37)[r]{$k_4$}
\Text(100,90)[r]{$W^+_\gamma$}
\Text(100,30)[r]{$W^-_\rho$}
\Photon(40,60)(80,30){2}{5}
\ArrowLine(60,75)(55,75)
\ArrowLine(60,45)(55,50)
\Text(120,60)[l]{$-i \;\frac{\displaystyle g^2 m_W^2 }{\displaystyle 
\Lambda^2} f_W \Big[ g^{\alpha\beta} g^{\gamma\rho}+ 
g^{\alpha\rho} g^{\beta\gamma}- 2g^{\alpha\gamma} g^{\beta\rho}\Big]$}
\end{picture}
\end{center}
\end{document}